\documentclass[
 reprint,
 amsmath,
 amssymb,
 aps,
 preprintnumbers,
 floatfix,
 nofootinbib
]{revtex4-1}

\usepackage{slashed}
\usepackage{dsfont}
\usepackage{amsmath}
\usepackage{graphicx}
\usepackage[caption=false]{subfig}
\usepackage{xspace}
\usepackage{color}
\usepackage{pictex}

\newcommand{\I}{\mathds{1}}
\newcommand{\Z}{\mathds{Z}}
\newcommand{\R}{\mathds{R}}
\newcommand{\SU}{\mathrm{SU}}
\newcommand{\su}{\mathfrak{su}}
\newcommand{\Zc}{\mathrm{Z}}
\newcommand{\U}{\mathrm{U}}

\newcommand{\muhat}{\hat{\mu}}
\newcommand{\nuhat}{\hat{\nu}}
\newcommand{\munu}{{\mu\nu}}
\newcommand{\eq}[1]{(\ref{{1}})}
\DeclareMathOperator{\sign}{sgn}

\DeclareMathOperator{\re}{Re}
\DeclareMathOperator{\Tr}{Tr}

\newcommand{\nn}{{\nonumber}}

\newcommand{\stapleupdag}{
\setlength{\unitlength}{14pt}
\begin{picture}(1,1)(0,0)
\linethickness{0.25pt}
\put(0,0){\circle*{0.15}}
\put(0,1){\vector(0,-1){1}}
\put(1,1){\vector(-1,0){1}}
\put(1,0){\vector(0,1){1}}
\put(1,0){\circle{0.14}}
\end{picture}}
\newcommand{\stapledowndag}{
\raisebox{-14pt}{
\setlength{\unitlength}{14pt}
\begin{picture}(1,1)(0,-1)
\linethickness{0.25pt}
\put(0,0){\circle*{0.15}}
\put(0,-1){\vector(0,1){1}}
\put(1,-1){\vector(-1,0){1}}
\put(1,0){\vector(0,-1){1}}
\put(1,0){\circle{0.14}}
\end{picture}}}

\newcommand{\rectup}{
\setlength{\unitlength}{14pt}
\raisebox{-14pt}{
\begin{picture}(1,2)(0,0)
\linethickness{0.25pt}
\put(0,0){\circle*{0.15}}
\put(1,0){\vector(0,1){2}}
\put(1,2){\vector(-1,0){1}}
\put(0,2){\vector(0,-1){2}}
\put(1,0){\circle{0.14}}
\end{picture}}}

\newcommand{\rectdown}{
\setlength{\unitlength}{14pt}
\raisebox{-14pt}{
\begin{picture}(1,2)(0,-2)
\linethickness{0.25pt}
\put(0,0){\circle*{0.15}}
\put(1,0){\vector(0,-1){2}}
\put(1,-2){\vector(-1,0){1}}
\put(0,-2){\vector(0,1){2}}
\put(1,0){\circle{0.14}}
\end{picture}}}

\newcommand{\rectrightmidleft}{
\setlength{\unitlength}{14pt}
\begin{picture}(2,1)(0,0)
\linethickness{0.25pt}
\put(0,0){\circle*{0.15}}
\put(1,0){\vector(1,0){1}}
\put(2,0){\vector(0,1){1}}
\put(2,1){\vector(-1,0){2}}
\put(0,1){\vector(0,-1){1}}
\put(1,0){\circle{0.14}}
\end{picture}}

\newcommand{\rectrightmidright}{
\setlength{\unitlength}{14pt}
\begin{picture}(2,1)(0,0)
\linethickness{0.25pt}
\put(1,0){\circle*{0.15}}
\put(0,0){\vector(1,0){1}}
\put(2,0){\vector(0,1){1}}
\put(2,1){\vector(-1,0){2}}
\put(0,1){\vector(0,-1){1}}
\put(2,0){\circle{0.14}}
\end{picture}}

\newcommand{\rectleftmidleft}{
\setlength{\unitlength}{14pt}
\raisebox{-14pt}{
\begin{picture}(2,1)(0,-1)
\linethickness{0.25pt}
\put(0,0){\circle*{0.15}}
\put(1,0){\vector(1,0){1}}
\put(2,0){\vector(0,-1){1}}
\put(2,-1){\vector(-1,0){2}}
\put(0,-1){\vector(0,1){1}}
\put(1,0){\circle{0.14}}
\end{picture}}}

\newcommand{\rectleftmidright}{
\setlength{\unitlength}{14pt}
\raisebox{-14pt}{
\begin{picture}(2,1)(0,-1)
\linethickness{0.25pt}
\put(1,0){\circle*{0.15}}
\put(0,0){\vector(1,0){1}}
\put(2,0){\vector(0,-1){1}}
\put(2,-1){\vector(-1,0){2}}
\put(0,-1){\vector(0,1){1}}
\put(2,0){\circle{0.14}}
\end{picture}}}

\begin{document}
\preprint{ADP-22-10/T1181}

\title{Smoothing algorithms for projected center-vortex gauge fields}

\author{Adam Virgili} \author{Waseem Kamleh} \author{Derek Leinweber}
\affiliation{Special Research Centre for the Subatomic Structure of
 	Matter, Department of Physics, University of Adelaide,
 	South Australia 5005, Australia.}

\begin{abstract}
 
    We study the application of $\SU(3)$ gauge field smoothing methods to $\Zc(3)$-projected center-vortex gauge fields.
    Due to the proportionality of the vortex links to the identity, naive applications of these methods are either ineffectual or limited in scope, containing subtle issues which are not obviously manifest.
    To overcome these issues we introduce \emph{centrifuge preconditioning}, a novel method applied prior to smoothing that rotates the links away from the center while preserving the fundamental structure of the vortex field.
    Additionally, the concept of \emph{vortex-preserved annealed smoothing} is formulated to ensure the smoothing procedure maintains the underlying vortex structure.
    The application of these new methods in the context of annealed smoothing applied to vortex fields is shown to successfully achieve the desired smoothness condition required for the study of more advanced operators.

\end{abstract}

\maketitle

\section{Introduction}
\label{sec:introduction}
Center vortices -- topological structures of the QCD vacuum ground-state fields -- are the prime candidate thought to underpin low-energy, nonperturbative QCD, in particular, its two key features -- confinement and dynamical chiral symmetry breaking~\cite{tHooft:1977nqb,tHooft:1979rtg,DelDebbio:1996lih,Faber:1997rp,DelDebbio:1998luz,Bertle:1999tw,Faber:1999gu,Engelhardt:1999fd,Engelhardt:1999xw,Engelhardt:2000wc,Bertle:2000qv,Langfeld:2001cz,Greensite:2003bk,Bruckmann:2003yd,Engelhardt:2003wm,Boyko:2006ic,Ilgenfritz:2007ua,Bornyakov:2007fz,OCais:2008kqh,Engelhardt:2010ft,Bowman:2010zr,OMalley:2011aa,Trewartha:2015ida,Greensite:2016pfc,Biddle:2018dtc,Spengler:2018dxt}.
On the lattice, center vortices are revealed by projecting each gauge link to an element of the center $\Zc(N)$ of $\SU(N)$ where
\begin{equation}
    \begin{aligned}
        \mathrm{Z}(N) &\equiv \{ g \in \SU(N)\, \vert\, gh = hg\, \forall\, h \in \SU(N) \} \\ 
        &= \{ e^{i 2 \pi n / N} \I\, \vert\, n \in \Z_N \}
    \end{aligned}
\end{equation}
is the set of elements in $\SU(N)$ which commute with every other element of the group.
To obtain the center-projected links, the gauge field is fixed to maximal center gauge (MCG) by choosing the gauge transform $U_\mu(x) \to U^G_\mu(x)$ which maximizes the functional~\cite{Langfeld:2003ev}
\begin{equation}
  \sum_{x,\mu}\left|{\Tr \, U^G_\mu(x)} \right|^2 \, ,
  \label{eq:MCG}
\end{equation}
as outlined in Refs.~\cite{Montero:1999by,Faber:1999sq}.
Each link is then projected to the nearest element of $\Zc(3)$, such that
\begin{equation}
    U^G_\mu(x) \to \mathcal{P}_{\Zc(3)} \left\{ U^G_\mu(x) \right\} \equiv Z_\mu(x) = e^{i\frac{2\pi}{3}n_\mu(x)}\I
    \label{eq:centre_project}
\end{equation}
where
\begin{equation}
    n_\mu(x) =
    \begin{cases}
        &\phantom{+}0, \text{ if } \arg\Tr\,U^G_\mu(x) \in \left(-\frac{\pi}{3}\,,+\frac{\pi}{3}\right) \,, \\
        &+1, \text{ if } \arg\Tr\,U^G_\mu(x) \in \left(+\frac{\pi}{3}\,,+\pi\right) \,, \\
        &-1, \text{ if } \arg\Tr\,U^G_\mu(x) \in \left(-\pi\,,-\frac{\pi}{3}\right) \,.
    \end{cases}
    \label{eq:centre_element}
\end{equation}
The projected links $Z_\mu(x)$ define a center-vortex configuration in MCG. The elementary plaquette $P_{\mu\nu}(x)$ is given by the product of links $U$ around a unit square, 
\begin{equation}
  P_{\mu\nu}(x) = U_\mu(x)\,U_\nu(x+\muhat)\,U_\mu^\dagger(x+\nuhat)\,U_\nu^\dagger(x).
\end{equation}
Center vortices are identified by the vortex flux through each vortex-projected plaquette, where
\begin{equation}
    \begin{aligned}
        P_{\mu\nu}(x) &= Z_\mu(x)\,Z_\nu(x+\muhat)\,Z_\mu^\dagger(x+\nuhat)\,Z_\nu^\dagger(x)\\ 
        &= e^{i\frac{2\pi}{3}p_\munu(x)}\I \, ,
    \end{aligned}
\end{equation}
corresponds to a vortex flux value $p_\munu(x) \in \left\{-1,0,1\right\}$. 
A plaquette with vortex flux $p_\munu(x)=\pm1$ is identified as pierced by a vortex with center charge $\pm1$.

Composed only of links which are elements of $\Zc(3)$, projected center-vortex gauge fields are rough, and naturally, violate the smoothness condition of the overlap Dirac operator~\cite{Hernandez:1998et,Adams:1999if,Neuberger:1999pz,Kamleh:2001ff}.
The vortex field must be smoothed.
Previous pure-gauge center-vortex studies using overlap fermions~\cite{Trewartha:2015nna,Trewartha:2017ive} have employed cooling to this end.
Whilst cooling suffices for vortex fields derived from pure gauge backgrounds, it is not ideal for smoothing fields derived from dynamical backgrounds.
In the dynamical case, an ideal smoothing algorithm would not only be analytical, but also preserve the underlying vortex structure. The constrained cooling algorithm~\cite{Langfeld:2010nm} gives an example of the importance of structure-preserving smoothing.

Whilst there has been a successful, novel approach to smoothing $\Zc(2)$ center vortex gauge fields~\cite{Hollwieser:2015koa}, this has not been generalized to $\Zc(3)$.
As such, this work focuses on applying existing $\SU(3)$ gauge field smoothing algorithms to $\Zc(3)$ center vortex gauge fields, with the goal of smoothing the vortex field such that the smoothness condition of the overlap Dirac operator is satisfied.

Section~\ref{sec:analytic_smoothing} examines the behavior of analytic smoothing methods on vortex fields; specifically stout smearing, gradient flow, and APE-smearing with analytic projection.
Section~\ref{sec:maxretr_proj} looks at update-based smoothing, in particular the MaxReTr reuniterization process and its application to cooling/annealing. 
Section~\ref{sec:centrifuge_preconditioning} offers a novel preconditioning method -- {\it{centrifuge preconditioning}} -- for $\Zc(3)$ center vortex gauge fields as a solution to the shortfalls and limitations of the traditional smoothing algorithms encountered in the previous sections.
Section~\ref{sec:centrifuge_preconditioned_smoothing} presents the results of the Wilson flow and annealed $U$-link smearing (AUS)~\cite{Bonnet:2000dc} applied to a centrifuge preconditioned gauge field.
Section~\ref{sec:vpas} introduces a vortex preservation step into the annealing process. 
Section~\ref{sec:comparison} compares three viable approaches to smoothing $\Zc(3)$ center vortices arrived at over the course of the work presented herein.
Section~\ref{sec:sum} summarizes the findings of this work.

\section{Analytic smoothing}
\label{sec:analytic_smoothing}

When smoothing Monte Carlo generated gauge fields, the use of analytic smoothing methods is inherently desirable. There are a number of such methods, the most commonly used being stout link smearing~\cite{Morningstar:2003gk} and the related gradient flow~\cite{Luscher:2009eq,Luscher:2010iy}. Through the use of a unitary projection method, it is also possible to apply APE-style blocking techniques whilst preserving analyticity~\cite{Kamleh:2004xk}. The differentiability of such smoothing methods is advantageous as it means that they can be applied within the molecular dynamics integration component of the Hybrid Monte Carlo algorithm~\cite{Duane:1987de,Kamleh:2004xk}. More importantly in the context of this work, the use of an analytic smoothing process implies that there is a parameterizable path within the gauge manifold that connects the original gauge links with the smoothed links.

The full $\SU(3)$ gauge group is described by 8 real parameters, whereas the center group $\Zc(3)$ consists of 3 discrete elements. The discrete nature of the center group presents significant challenges when attempting to apply standard smoothing techniques, as we demonstrate below. First we define a quantity which is relevant to all the methods considered herein, namely the sum of the \emph{staples} orthogonal to a link $U_\mu(x),$
\begin{multline}
    \Sigma_\mu(x) = \sum_{\nu \ne \mu} \left[U_\nu(x)\,U_\mu(x + \nuhat)\,U^\dagger_\nu(x + \muhat)\right. \\ 
     \left. + U^\dagger_\nu(x - \nuhat)\,U_\mu(x - \nuhat)\,U_\nu(x - \nuhat + \muhat)\right] \, . 
    \label{eq:plaq_staples}
\end{multline}
Related to the above, we also introduce the sum of the corresponding plaquettes (by closing the path of the staples via link multiplication) as
\begin{equation}
    \Omega_\mu(x) = \Sigma_\mu(x)\,U^\dagger_\mu(x) \,.
\end{equation}
  
\subsection{Stout link smearing}
\label{sec:stout smearing}

Stout link smearing~\cite{Morningstar:2003gk} provides the simplest case to show the difficulties of analytic smearing of center-vortex projected fields. A single iteration of stout link smearing with isotropic smearing parameter $\rho$ is defined by
\begin{equation}
  \tilde{U}_\mu(x) = \exp\left( \rho\, Q_\mu(x) \right) U_\mu(x),
  \label{eq:stout}
\end{equation}
where $Q_\mu(x)$ is the traceless anti-Hermitian projection of $\Omega_\mu(x)$ onto the Lie algebra $\su(3)$,
\begin{equation}
  Q_\mu(x) = \frac{1}{2}\left[\Omega_\mu(x) - \Omega^\dagger_\mu(x)\right] - \frac{1}{6}\Tr\left[\Omega_\mu(x) - \Omega^\dagger_\mu(x)\right] \,.
  \label{eq:generator}
\end{equation}
In the case of a vortex field, as each of the center elements $Z_\mu(x) \propto \I$ is proportional to the identity matrix, we can parameterize any sum of vortex link paths as $re^{i\theta}\I$ where $r, \theta \in \R$. Consequently, we have that
\begin{align}
  Q_\mu(x) &= \frac{1}{2} \left[ re^{i\theta} - re^{-i\theta} \right]\I - \frac{1}{6}\Tr\left[ (re^{i\theta} - re^{-i\theta})\I \right]\I \nonumber \\
           &= 0.
\end{align}
As $Q_\mu(x)$ vanishes when derived from a center vortex field, it immediately follows from Eq.~\ref{eq:stout} that $\tilde{U}_\mu(x) = Z_\mu(x).$ That is stout link smearing leaves the vortex field unchanged. This result remains true in the presence of a gauge transformation $G(x)$, as we have in general that
\begin{align}
    \Sigma_\mu(x) &\to \Sigma^G_\mu(x) = G(x)\,\Sigma_\mu(x)\,G^\dagger(x+\muhat) \,, \\
    \Omega_\mu(x) &\to \Omega^G_\mu(x) = G(x)\,\Omega_\mu(x)\,G^\dagger(x) \,.
\end{align}
In the case of a vortex field, as the center group commutes with all other elements by definition, then we have that
\begin{equation}
    \Omega^G_\mu(x) \to re^{i\theta}\,G(x)\,G^\dagger(x) = re^{i\theta}\I \,,
\end{equation}
proving that a gauge transformed center vortex field is also invariant under stout smearing.

\subsection{Gradient flow}
\label{sec:gradient_flow}

The gradient flow~\cite{Luscher:2009eq,Luscher:2010iy} is defined by the equations
\begin{align}
    \frac{d}{d\tau}U_\mu(x;\tau) &= Q_\mu(x)[U(\tau)]\,U_\mu(x;\tau) \,,\\ 
    U_\mu(x;0) &= U_\mu(x) \,,
\end{align}
where $\tau$ is dimensionless {\it{flow time}} and $Q_\mu(x)[U(\tau)] \in \su(3)$ is the generator of the infinitesimal field transformation
\begin{equation}
    U \to U + \epsilon\, Q(U) \, U + \mathcal{O}(\epsilon^2) \,.
\end{equation}
In particular, the {\it{Wilson flow}} is generated by
\begin{equation}
    Q_\mu(x)[U] = T^a\,\partial^a_{x,\mu} \sum_{x,\mu \ne \nu} \Tr\left[P_\munu(x)[U]\right]
\end{equation}
where $T^a$ are the generators of $\SU(3)$ (see Appendix~\ref{sec:generatorsSU3}) and
\begin{align}
    \partial^a_{x,\mu}f(U) &= \left.\frac{d}{ds} f(e^{sX^a}\,U) \right|_{s=0} \,, \\
    X^a(y,\nu) &= 
    \begin{cases}
        T^a \text{ if } (y,\nu) = (x,\mu) \\
        0 \text{ otherwise. }
    \end{cases} 
\end{align}
The explicit formula for the generator $Q_\mu(x)[U]$ is identical to that for stout smearing given in Eq.~\ref{eq:generator}. In fact, noting that
\begin{equation}
\lim_{n\to\infty} (\I + \epsilon\, Q)^n U = \exp(\epsilon\, Q)U
\end{equation}
we can map $\epsilon \to \rho$ and see that the stout-smeared link is the finite transformation generated by the Wilson flow process for sufficiently small smearing parameters.

It trivially follows that for a center vortex field $U_\mu(x;0) = Z_\mu(x)$ we have
\begin{equation}
    \frac{d}{d\tau}U_\mu(x;\tau) = 0 \ \ \forall \ \ x, \ \ \mu, \ \ \tau \,.
\end{equation}
Hence, independent of the initial gauge or the integration method, the Wilson flow of a center vortex gauge field is invariant.

\subsection{APE smearing with analytic projection}
\label{sec:ape_analytic}

We now consider APE smearing~\cite{Falcioni:1984ei,APE:1987ehd} with the analytic projection method defined in Ref.~\cite{Kamleh:2004xk}, which we refer to as \emph{unit circle projection}.
The APE smearing process starts with a blocking step, where the original link $U_\mu(x)$ is mixed with the sum of the staples in proportion to the smearing parameter $\alpha$ to define the blocked matrix
\begin{equation}
  V\left[U_\mu(x)\right] \equiv V_\mu(x) = (1 - \alpha) \, U_\mu(x) + \frac{\alpha}{6}\,\Sigma_\mu(x) \,.
  \label{eq:V}
\end{equation}
This construction of $V_\mu(x)$ is gauge equivariant, which is to say, under a gauge transformation 
\begin{equation}
    U_\mu(x) \to U^G_\mu(x) = G(x)\,U_\mu(x)\,G^\dagger(x+\muhat)
\end{equation}
that
\begin{equation}
    V\left[U^G_\mu(x)\right] = G(x)\,V\left[U_\mu(x)\right]\,G^\dagger(x+\muhat) \,.
\end{equation}
Setting $U^{(0)}_\mu(x) = U_\mu(x),$ the APE smearing update is then defined by
\begin{equation}
    U^{(n)}_\mu(x) \to U^{(n+1)}_\mu(x) = \mathcal{P} \left\{ V^{(n)}_\mu(x) \right\} \,,
\end{equation}
where the blocked matrix $V^{(n)}_\mu(x) \equiv V[U^{(n)}_\mu(x)] \notin \SU(3)$ must be returned back to the gauge group. This may be performed in an analytic manner by first performing a unitary projection,
\begin{equation}
    W_\mu(x) = V_\mu(x)\,\frac{1}{\sqrt{V^\dagger_\mu(x)\,V_\mu(x)}} \,,
    \label{eq:W}
\end{equation}
such that the eigenvalues of $W$ lie on the unit circle. The final step in the unit circle projection is multiplying by the appropriate phase in order to return to $\SU(3)$,
\begin{equation}
    \mathcal{P}_{\rm ucp} \left\{ V_\mu(x) \right\} = \frac{1}{\sqrt[3]{\det W_\mu(x)}}\,W_\mu(x) \, .
\end{equation}
As shown in Ref.~\cite{Kamleh:2004xk}, the unit circle projection is gauge equivariant such that the smeared links share the same gauge transformation properties as the original link,
\begin{equation}
    U^{(n)}_\mu(x) \to G(x)\,U^{(n)}_\mu(x)\,G^\dagger(x+\muhat).
\end{equation}
In standard APE smearing, the staples term $\Sigma_\mu(x)$ is defined as per equation~(\ref{eq:plaq_staples}), but other choices are possible, in particular the over-improvement formalism~\cite{GarciaPerez:1993lic,Bonnet:2001rc,Moran:2008ra} outlined in Eq. (\ref{eq:over-improvement}).
For the purposes of the following discussion, we generalize $\Sigma_\mu(x)$ to sum over any combination of operators constructed from paths originating at lattice site $x$ and terminating at $x+\muhat$.

When APE smearing is applied to a center vortex gauge field in arbitrary gauge (noting that $Z_\mu(x) \propto \I$), we can use gauge equivariance to write
\begin{equation} 
    V^{(0)}_\mu(x) = re^{i\theta}\,G(x)\,G^\dagger(x+\muhat) \,.
    \label{eq:V_VO}
\end{equation}
Applying the unitary projection in Eq.~\ref{eq:W} gives
\begin{align}
  W_\mu(x) &= re^{i\theta}\,G(x)\,G^\dagger(x+\muhat)\,\frac{\I}{\sqrt{r^2}} \nonumber \\ 
  &= e^{i\theta}\,G(x)\,G^\dagger(x+\muhat) \,.
\end{align}
Noting that $\det W_\mu(x) = e^{i3\theta},$ we have that
\begin{equation}
    \sqrt[3]{\det W_\mu(x)} = e^{i\left(\theta + \frac{2k\pi}{3}\right)} \,, 
\end{equation}
where $k \in \{0,1,2\} $ is chosen to correspond to the principal cube root, i.e. such that
\begin{equation}
    -\frac{\pi}{3} < \theta + \frac{2k\pi}{3} < \frac{\pi}{3} \, .
\end{equation}
Hence, the projected link is given by
\begin{align}
  Z^{(1)}_\mu(x) &= \frac{1}{e^{i\left(\theta + \frac{2k\pi}{3}\right)}}\,e^{i\theta}\,G(x)\,G^\dagger(x+\muhat) \nonumber \\
  &= e^{-i\frac{2k\pi}{3}}\,G(x)\,G^\dagger(x+\muhat) \,,
\end{align}
where $e^{-i\frac{2k\pi}{3}}\I \in \Zc(3)$.

The key result here is that it is only possible to project to another element of $\Zc(3).$ That is, applying APE smearing with unit circle projection to a vortex link
results in either the original link remaining unchanged, or selecting a completely different center element and thereby radically altering the vortex structure such that it no longer resembles its original form.
As the method is gauge equivariant, this is true regardless of whether we are in maximal center gauge or not.

\section{Update-based smoothing}
\label{sec:maxretr_proj}

Having determined that none of the analytic smearing techniques considered above can smoothly deform a vortex field away from the center group, we consider an nonanalytic alternative. Specifically, we examine APE-style blocking coupled with the update-based reuniterization method maximizing the real part of the trace (MaxReTr) of the plaquette. This process is based on the Cabibbo-Marinari pseudo-heat-bath algorithm~\cite{Cabibbo:1982zn} which iteratively updates a candidate $SU(N)$ matrix $U_\mu(x)$ to maximize the following,
\begin{equation}
    \max\re\Tr\left[ U_\mu(x)\,V^\dagger_\mu(x) \right],
    \label{eq:maxReTr}
\end{equation}
where $V_\mu(x)$ is the sum of link paths defined in Eq.~(\ref{eq:V}). MaxReTr reuniterization is fundamentally connected with cooling~\cite{Bonnet:2001rc}, as if we set the smearing fraction $\alpha=1,$ then we have $V_\mu(x) \propto \Sigma_\mu(x)$ and then the MaxReTr update selects the link which minimizes the local action in a way which does not depend on the original link.

Due to the nonanalytic nature of the MaxReTr update process, it is able to shift the vortex fields away from the center group in a way that the differentiable smoothing methods above cannot. It will prove useful to review the specific details of the MaxReTr method as applied to $\SU(3),$ which involves iterating over $SU(2)$ subgroups in order to achieve the optimization specified by Eq.~(\ref{eq:maxReTr}).
First, define the matrix $L^1$ by
\begin{equation}
    L^1 = U_\mu(x)\,V^\dagger_\mu(x) \,.
    \label{eq:def_L1}
\end{equation}
From $L^1$, another matrix, $T_1 \in \SU(2) \subset \SU(3)$, given by
\begin{equation}
    T_1 = \frac{1}{2}
    \begin{bmatrix}
        L^1_{11} + (L^1_{22})^* & L^1_{12} - (L^1_{21})^* & 0 \\
        L^1_{21} - (L^1_{12})^* & (L^1_{11})^* + L^1_{22} & 0 \\
    0 & 0 & 2
  \end{bmatrix}
    \,,
    \label{eq:def_T1}
\end{equation}
is constructed, where $L^1_{ij}$ is element $(i,j)$ of $L^1.$ Setting $k = \sqrt{\det T_1},$ this matrix is cooled such that
\begin{equation}
    \big[T_1^c\big]_{2\times 2} = \frac{1}{k}\,\big[T^\dagger_1\big]_{2\times 2}\in\SU(2)\,,
\end{equation}
within the embedded $2\times 2$ subgroup, and the full matrix $T^c_1$ is the embedding of the resulting submatrix into $\SU(3).$
The original link is then updated by
\begin{equation}
    U_\mu(x) \to U'_\mu(x) = T^c_1 U_\mu(x) \,,
\end{equation}
This process is typically repeated for the other two diagonal $\SU(2)$ subgroups which together comprehensively cover $\SU(3)$, such that
\begin{align}
    L^2 &= U'_\mu(x)V^\dagger_\mu(x) = T^c_1U_\mu(x)V^\dagger_\mu(x) \,, \\
    L^3 &= U''_\mu(x)V^\dagger_\mu(x) = T^c_2T^c_1U_\mu(x)V^\dagger_\mu(x) \,,
\end{align}
and
\begin{align}
    T_2 = \frac{1}{2}
    &\begin{bmatrix}
        2 & 0 & 0 \\
        0 & L^2_{22} + (L^2_{33})^* & L^2_{23} - (L^2_{32})^* \\
        0 & L^2_{32} - (L^2_{23})^* & (L^2_{22})^* + L^2_{33}
    \end{bmatrix}
    \,,\\
    T_3 = \frac{1}{2}
    &\begin{bmatrix}
        L^3_{11} + (L^3_{33})^*  & 0 & L^3_{13} - (L^3_{31})^*  \\
        0 & 2 & 0 \\
        L^3_{31} - (L^3_{13})^* & 0 & (L^3_{11})^* + L^3_{33}
    \end{bmatrix}
    \,.
\end{align}

The updated link $U^{(1)}_\mu(x)$ after one loop over the $\SU(2)$ subgroups is given by
\begin{equation}
    T^c_3\,T^c_2\,T^c_1\,U_\mu(x) \,.
\end{equation}
In principle, one loop over the subgroups is considered sufficient with regard to approaching the maximum defined by Eq.~(\ref{eq:maxReTr}). Here we choose to perform three iterations over the subgroups as multiple loops provide an advantage in converging to the optimal link~\cite{Bonnet:2001rc}.

Let us now explore how the MaxReTr reuniterization algorithm applies to a center vortex gauge field which has undergone an arbitrary gauge transformation
\begin{equation}
    Z_\mu(x) \to G(x)\,Z_\mu(x)\,G^\dagger(x+\muhat) \,.
\end{equation}
Using the gauge equivariance of $V_\mu(x)$, $L^1$ has the gauge invariant form
\begin{align}
  L^1 &= G(x)\,Z_\mu(x)\,G^\dagger(x+\muhat)\,G(x+\muhat)\,V^\dagger_\mu(x)\,G^\dagger(x) \nn \\
  &= G(x)\,re^{i\left(\frac{2\pi n}{3} - \theta\right)}\,G^\dagger(x) \nn \\
  &\equiv re^{i\phi}\,\I \,,
  \label{eq:L1_invariance}
\end{align}
where $V_\mu(x) = re^{i\theta}$ is in MCG and we have defined $\phi \equiv \frac{2\pi n}{3} - \theta$ for $Z_\mu(x) = e^{i\frac{2\pi n}{3}}\,\I$ also in MCG.
Hence, $T_1$ is given by
\begin{align}
  T_1 &=\frac{1}{2} 
  \begin{bmatrix}
    r(e^{i\phi} + e^{-i\phi}) & 0 & 0 \\
    0 & r(e^{i\phi} + e^{-i\phi}) & 0 \\
    0 & 0 & 2
  \end{bmatrix} \nn \\
  &=\phantom{\frac{1}{2}} 
  \begin{bmatrix}
    r \cos \phi & 0 & 0 \\
    0 & r \cos \phi & 0 \\
    0 & 0 & 1
  \end{bmatrix}\,.
\end{align}
It follows then, that
\begin{equation}
    k = \sqrt{\det T_1} = \sqrt{r^2\cos^2\phi} = \vert r\cos\phi\vert \,,
\end{equation}
and
\begin{equation}
    T^c_1 = 
    \begin{bmatrix}
        \sign\left({r \cos \phi}\right) & 0 & 0 \\
        0 & \sign\left({r \cos \phi}\right) & 0 \\
        0 & 0 & 1
    \end{bmatrix}
    \,.
\end{equation}
If $r\cos\phi > 0 \implies T^c_1 = \I$ and
\begin{equation}
    Z'_\mu(x) = T^c_1\,Z_\mu(x) = Z_\mu(x) \,.
\end{equation}
It is straightforward to see that $T^c_1 = \I \implies T^c_i = \I \ \ \forall \ \ i$, and hence
\begin{equation}
    Z^{(1)}_\mu(x) = Z_\mu(x) \,.
\end{equation}
By induction,
\begin{equation}
    Z^{(n)}_\mu(x) = Z_\mu(x) \ \ \forall \ \ n \,,
\end{equation}
and the center vortex field is unchanged.

As such, in order to perturb the vortex field we require $r\cos\phi < 0$, or equivalently, $\left| \phi \right| > \frac{\pi}{2}$.
This necessarily places a condition on the smearing parameter $\alpha$.
We consider this condition within the context of the over-improvement formalism~\cite{Moran:2008ra}, for which the staples term is given by the diagrammatic equation
\begin{widetext}
\begin{equation}
\Sigma^\dagger_\mu(x) = \sum_{\nu \ne \mu} \frac{5-2\epsilon}{3} \left( \stapleupdag + \stapledowndag \right) + \frac{\epsilon-1}{12u^2_0} \left(  \rectup + \rectdown\ + \rectleftmidleft + \rectleftmidright + \rectrightmidleft + \rectrightmidright\ \right),
\label{eq:over-improvement}
\end{equation}
\end{widetext}
where the solid dot represents the point $x,$ the open dot represents the point $x+\muhat,$ and the links in the positive orthogonal direction $\nuhat$ are shown as pointing vertically up the page. Note also that we have illustrated the link paths as oriented for the Hermitian conjugate $\Sigma^\dagger_\mu(x)$ which enters into Eq.~(\ref{eq:maxReTr}).

The over-improvement formalism encapsulates standard APE smearing at over-improvement term $\epsilon = 1$. 
To ensure $r\cos\phi < 0$ we require that
\begin{equation}
    \alpha > \alpha_\text{min} = \frac{-6}{2\epsilon - 11 + \frac{3}{2}\left(\frac{\epsilon-1}{u_0^2}\right)}\,,
    \label{eq:alpha_condition_ape}
\end{equation}
for $\epsilon \in \left[-\frac{5}{2},1\right]$, where $u_0$ is the mean link.
See Appendix~\ref{sec:alpha_min_derivation} for a derivation.

\begin{figure}[t]
    \subfloat[$\epsilon=-0.25$, $\alpha=0.7$]{\includegraphics[width=0.9\columnwidth]{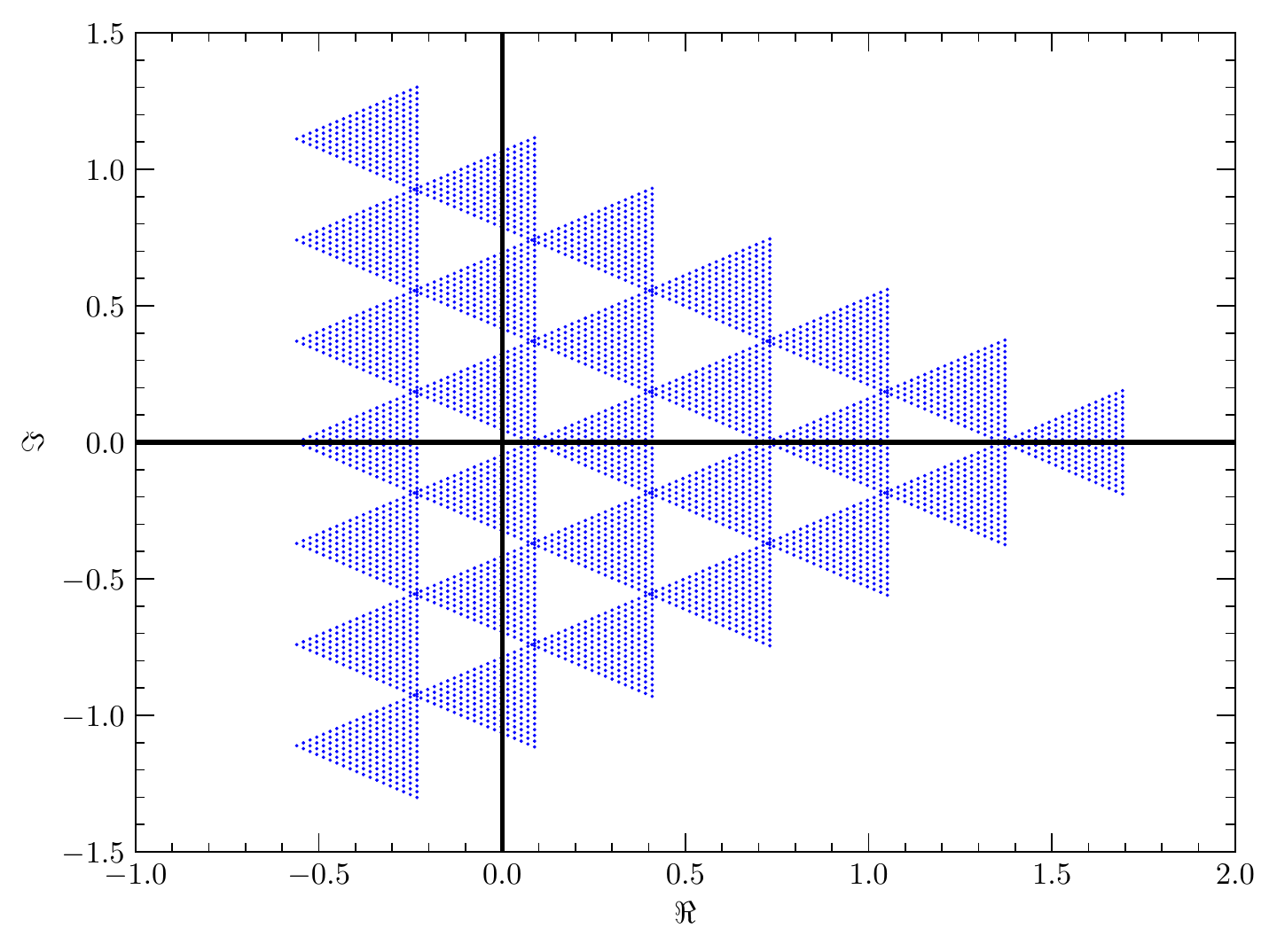}}
    \\
    \subfloat[$\epsilon=-0.25$, $\alpha=0.4$]{\includegraphics[width=0.9\columnwidth]{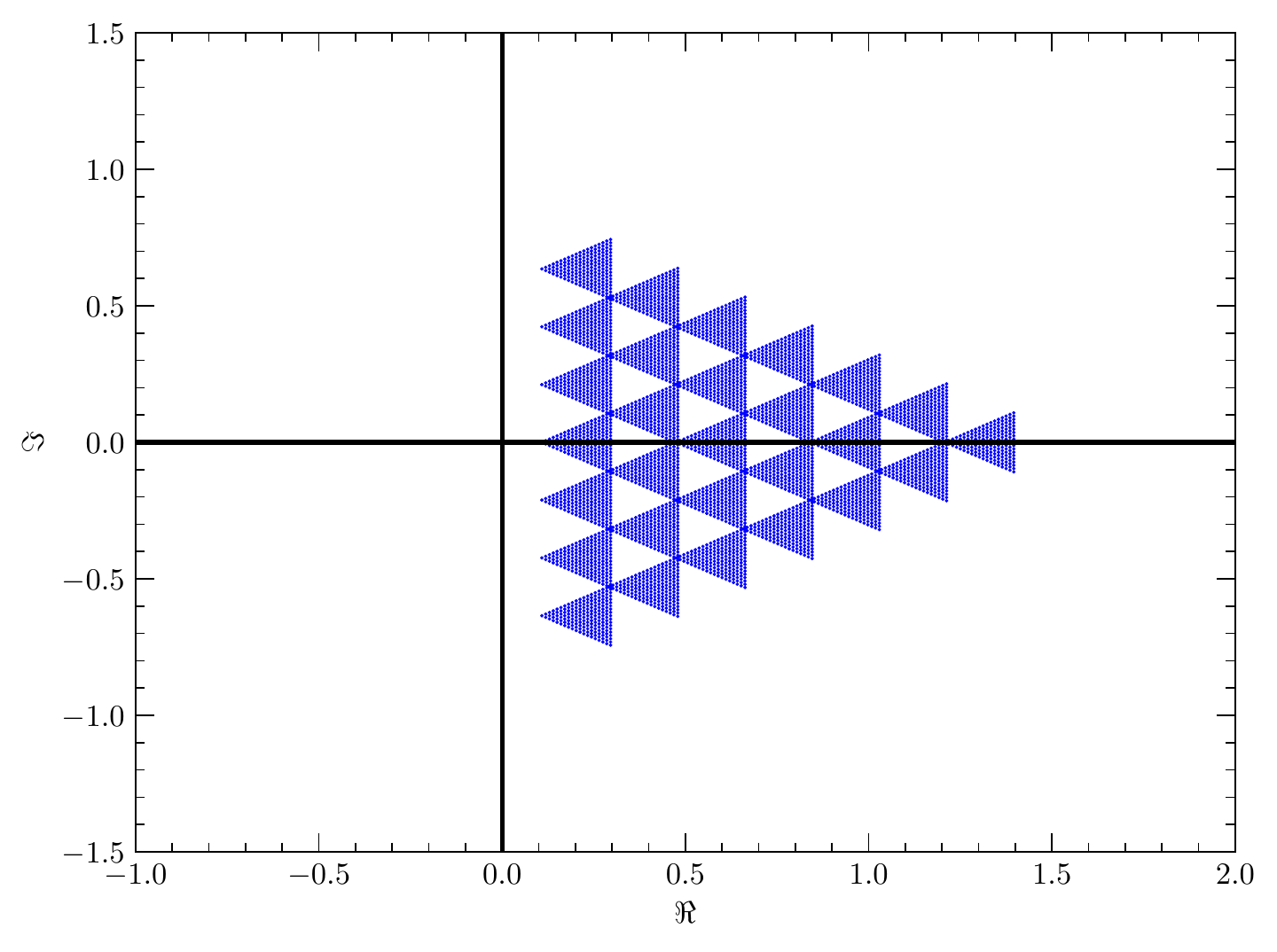}}
    \caption{The complex plane showing all possible values of $re^{i\phi}$ for over-improvement term $\epsilon=-0.25$, at smearing parameters (a) $\alpha=0.7$ and (b) $\alpha=0.4$.}
    \label{fig:L1}
\end{figure}

In Fig.~\ref{fig:L1}, all possible values of $re^{i\phi}$ originating from a center-vortex configuration for $\epsilon=-0.25$ and $u_0=1$, at $\alpha=0.4$ and $\alpha=0.7$, respectively, are plotted on the complex plane.
At $\alpha=0.7$ there are many combinations of links which yield $\left| \phi \right| > \frac{\pi}{2}$, but none at $\alpha=0.4$.
In fact, from equation~(\ref{eq:alpha_condition_ape}), $\alpha_\text{min}(\epsilon=-0.25,u_0=1) \approx 0.4486$.
Of course, this is the limit to have {\it{any}} combination of links yield $\left| \phi \right| > \frac{\pi}{2}$.
In a practical sense, we require something more like $\alpha > 0.6$ to achieve effective smearing.

\begin{figure}[t]
    \includegraphics[width=0.9\columnwidth]{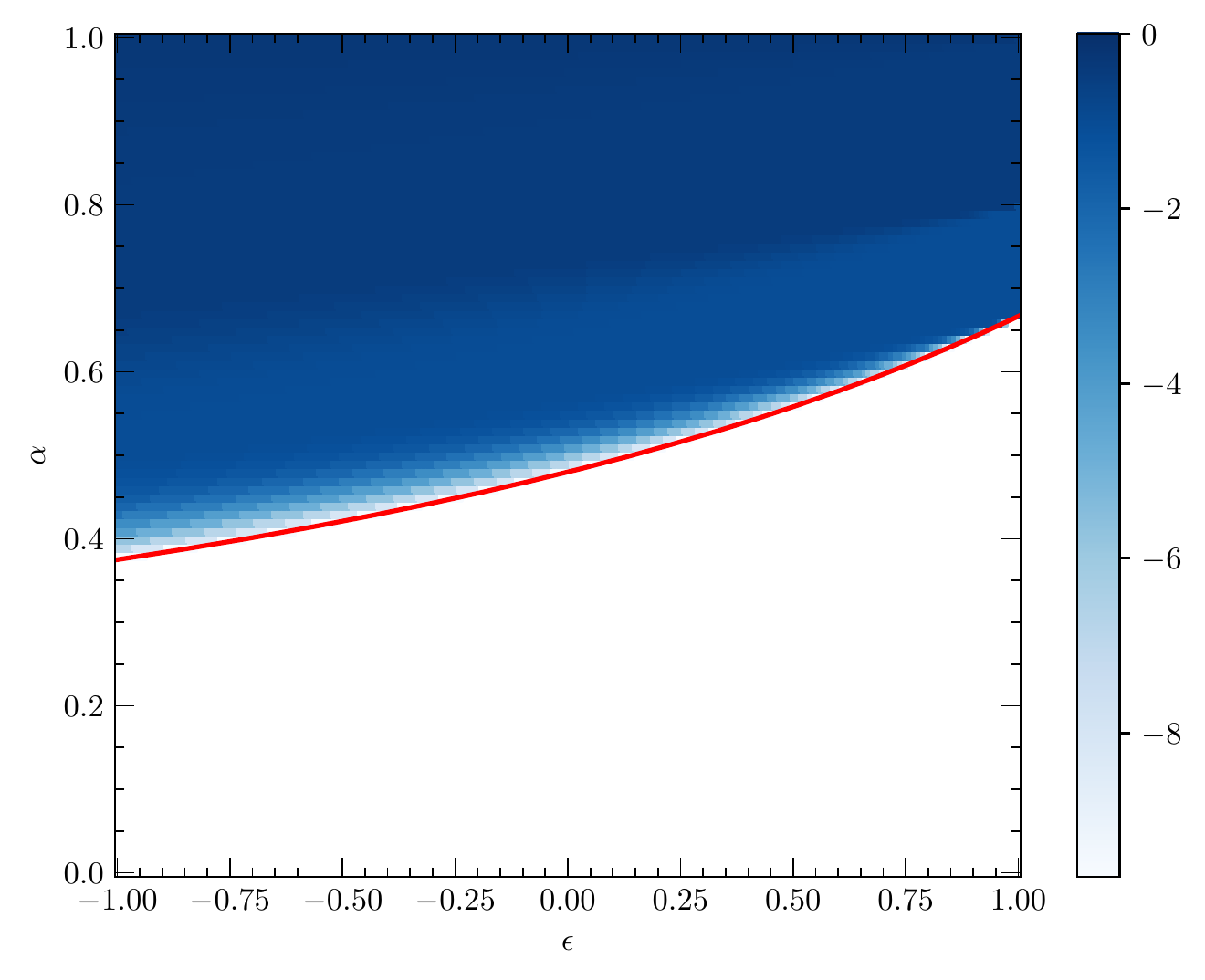}
    \caption{$\log_{10}(p_\text{in})$ as a function of $\epsilon$ and $\alpha,$ at $u_0=1$ fixed in the limit $\beta \to 0$. The shaded region illustrates the admissible values of $\alpha$ and $\epsilon$ where the proportion of acceptable link combinations, $p_\text{in}$, exceeds zero. The red line is $\alpha_\text{min}(\epsilon)$.}
    \label{fig:p_in}
\end{figure}

Fig.~\ref{fig:p_in} presents the proportion $p_\text{in}$ of link combinations for which $\left| \phi \right| > \frac{\pi}{2}$ in the $\beta \to 0$ limit where each link in the construction of $L^1$ has an equal probability to be one of the center phase elements.
Each combination is weighted by its multiplicity. 
These do not reflect the true probabilities which would be encountered on an actual $\Zc(3)$ center vortex gauge field, but suffice for demonstrative purposes.

\subsection{Cooling/annealing}
\label{sec:note_on_cooling}

While smearing algorithms update all links simultaneously, smoothing via a cooling or annealing process updates each link individually. These updates can be done in parallel with appropriate masking so as to preserve the validity of the cooling or annealing process~\cite{Bonnet:2000db}. The Wilson flow can be considered as an annealed version of stout link smearing with small smearing parameter. \emph{Annealed U-link smearing} (AUS)~\cite{Bonnet:2000dc} is similarly related to APE smearing in that the update process that uses APE-style blocking and reuniterization, but applied to individual links rather than all links. In particular, at $\alpha=1.0$ the form of AUS with MaxReTr reuniterization at the individual link level reduces to that of cooling~\cite{Berg:1981nw,Teper:1985rb,Teper:1985gi,Teper:1985ek,Ilgenfritz:1985dz} -- up to choice of operators in the staples term.

Our analysis with regard to vortex smoothing above extends to the annealed form of the various methods, and also to cooling in the special case of $\alpha=1.0$. With regard to cooling it should be noted that it is possible to encounter some numerical issues when smoothing center vortex fields.
From equations (\ref{eq:def_L1}) and (\ref{eq:L1_invariance}), we can write $L^1$ in arbitrary gauge as
\begin{equation}
    L^1 = (1-\alpha) + \frac{\alpha}{6}\,Z_\mu(x)\,\Sigma^\dagger_\mu(x)
    \label{eq:L_gauge_invariant}
\end{equation}
which reduces to 
\begin{equation}
    L^1 = \frac{1}{6}\,Z_\mu(x)\,\Sigma^\dagger_\mu(x)
\end{equation}
for cooling ($\alpha=1.0$), where $Z_\mu(x)$ and $\Sigma_\mu(x)$ are in MCG, i.e. they are proportional to $\I$.

Since $\Sigma_\mu(x)$ is a sum of elements of $\Zc(3)$, each multiplied by some real factor, there exists combinations of links for which $\Sigma_\mu(x) = is\I$ and $s \in \R$, which is to say that the nonzero (diagonal) elements of $\Sigma_\mu(x)$ are purely imaginary.
For example, in standard Wilson cooling where the 6 operators comprising the staples term are split two-to-four between $\I$ and $e^{\pm i\frac{2\pi}{3}}$ are examples of such combinations.
In these cases the staples term is given by
\begin{equation}
    \Sigma_\mu(x) = 2\I + 4e^{\pm i\frac{2\pi}{3}} = is \approx \pm i\,3.464... \,.
\end{equation}
Without loss of generality, choose $Z_\mu(x) = \I$. Then
\begin{align}
  L^1 &= Z_\mu(x)\,V^\dagger_\mu(x) \nn \\
  &= 
  \begin{bmatrix}
    -is & 0 & 0 \\
    0 & -is & 0 \\
    0 & 0 & -is
  \end{bmatrix}
  \,.
\end{align}
Constructing $T_1$ according to equation~(\ref{eq:def_T1}),
\begin{align}
  T_1 &= \frac{1}{2}
  \begin{bmatrix}
    -is + is & 0 & 0 \\
    0 & +is - is & 0 \\
    0 & 0 & 2
  \end{bmatrix} \nn \\
  &= \phantom{\frac{1}{2}}
  \begin{bmatrix}
    0 & 0 & 0 \\
    0 & 0 & 0 \\
    0 & 0 & 1
  \end{bmatrix}
\end{align}
which implies $k=\sqrt{\det T_1}=0$ and thus
\begin{equation}
    \big[T_1^c\big]_{2\times 2} = \frac{1}{k}\,\big[T^\dagger_1\big]_{2\times 2}\,,
\end{equation}
is undefined.

In practice, the diagonal elements of $T_1$ are not precisely $0$ due to floating-point artifacts.
From these artifacts, the algorithm is able to generate an essentially random $\SU(3)$ link without breaking or resulting in any obvious errors.
This is apparent in Table~\ref{t:VO.vpa}, where at $\alpha=1.0$, the number of links for which $Z^{(1)}_\mu(x) \ne Z_\mu(x)$, $n_\text{diff}$, is greater than $n_\text{in}$, the number of links for which $\vert \phi \vert > \frac{\pi}{2}$.
Whilst this situation is rare (12 of 1,280,000 links in Table~\ref{t:VO.vpa}), the link, generated essentially from random noise, contaminates neighboring links on the next iteration when it contributes to the staples term.
The noise contamination continues to propagate throughout the lattice after each successive sweep, as the contaminated neighbors then contaminate their neighbors.

\section{Centrifuge preconditioning}
\label{sec:centrifuge_preconditioning}

The issues and limitations, outlined in previous sections, which arise when applying traditional smoothing methods to center vortex gauge fields, are all, at root, due to the proportionality of the links to the identity.
As such, we introduce a method to break this symmetry without altering the fundamental vortex structure of the field. The key idea is to rotate the vortex links away from the center elements before applying smoothing, and hence we call this new method \emph{centrifuge preconditioning}.

We start with the original center vortex gauge field in MCG and denote
\begin{equation}
    Z_\mu(x) = 
    \begin{bmatrix}
        e^{i\lambda^1_\mu(x)} & 0 & 0 \\
        0 & e^{i\lambda^2_\mu(x)} & 0 \\
        0 & 0 & e^{i\lambda^3_\mu(x)} \\
    \end{bmatrix}
    \,,
\end{equation}
where initially the diagonal entries $\lambda^i_\mu(x) = \lambda_\mu(x)$ are all equal. Noting that we are now within the diagonal subgroup of $\SU(3),$ which is isomorphic to $\U(1)\times \U(1)\times \Zc(3)$,
we can work with the phases directly in the non-compact representation. Define the staple phase as
\begin{multline}
    \sigma_\mu(x) = \frac{1}{6}\sum_{\nu \ne \mu} \left[ \lambda_\nu(x) + \lambda_\mu(x+\nuhat) - \lambda_\nu(x+\muhat) \right. \\
    \left. - \lambda_\nu(x-\nuhat) + \lambda_\mu(x-\nuhat) + \lambda_\nu(x-\nuhat+\muhat) \right]\,.
\end{multline}
A pair of indices $(j,k) \in \left\{(1,2),\,(2,3),\,(3,1)\right\}$ are selected randomly for each link, and then the corresponding phases of each original link are updated according to
\begin{align}
    \label{eq:lambda_j}
    \lambda^j_\mu(x) &\to (1-\omega)\,\lambda_\mu(x) + \omega\,\sigma_\mu(x) \,,\\
    \label{eq:lambda_k}
    \lambda^k_\mu(x) &\to (1+\omega)\,\lambda_\mu(x) - \omega\,\sigma_\mu(x) \,,
\end{align}
where $\omega \in \R$ specifies the \emph{centrifugal rotation angle,} noting that the centrifuge update above corresponds to a phase rotation by $\mp\omega(\lambda - \sigma).$
This leaves the sum of the three phases invariant. Hence, as the sum of the three phases of each center element is distinct,
\begin{equation}
\sum_j \lambda^j_\mu(x) = n\,2\pi,\quad n \in \{-1,0,1\},
\end{equation}
after centrifuge preconditioning it is possible to uniquely identify the original center element by this sum.

\subsection{Preservation of vortex structure}

Recall from equations (\ref{eq:centre_project}) and (\ref{eq:centre_element}) that the center-vortex links are obtained by projecting the untouched link in maximal center gauge $U^{G}_\mu(x)$ to the center element with phase nearest to $\arg\Tr\,U^G_\mu(x)$.
Since we seek to break the diagonal symmetry of the center vortex links in such a way that preserves the underlying vortex structure, we restrict $\omega$ in equations (\ref{eq:lambda_j}) and (\ref{eq:lambda_k}) such that $\arg\Tr\,Z_\mu(x) = 2\pi n_\mu(x) / 3 $ is preserved.

Let $Z'_\mu(x)$ denote the preconditioned center vortex link with updated phases $\lambda^j_\mu(x)$ and $\lambda^k_\mu(x)$.
It is simple to see that
\begin{equation}
    \Tr \left[ Z'_\mu(x) \right] = e^{i\lambda^j_\mu(x)} + e^{i\lambda^k_\mu(x)} + e^{i\frac{2 \pi n}{3}} \,.
\end{equation}
We then define
\begin{equation}
    \Lambda^{\pm} = \frac{1}{2}\left( \lambda^j_\mu(x) \pm \lambda^k_\mu(x) \right).
\end{equation}
Utilizing polar form $e^{iA} = \cos A + i\sin A$, and the following trigonometric properties,
\begin{align}
    \cos A + \cos B = 2\cos\frac{A+B}{2}\cos\frac{A-B}{2} \,, \\
    \sin A + \sin B = 2\sin\frac{A+B}{2}\cos\frac{A-B}{2} \,,
\end{align}
we obtain
\begin{equation}
    \begin{aligned}
        e^{i\lambda^j_\mu(x)} + e^{i\lambda^k_\mu(x)} &= 2\cos\Lambda^+\cos\Lambda^- + i2\sin\Lambda^+\cos\Lambda^- \\
        &= 2\cos\Lambda^-\left(\cos\Lambda^+ - i\sin\Lambda^+\right) \\
        &= 2\cos\Lambda^-e^{i\frac{2\pi n}{3}} \,.
    \end{aligned}
\end{equation}
Hence,
\begin{equation}
    \Tr \left[ Z'_\mu(x) \right] = \left( 2\cos\Lambda^- + 1 \right) e^{n\frac{2\pi i}{3}} 
\end{equation}
and the phase of the trace is preserved,
\begin{equation}
    \arg\Tr\left[ Z_\mu(x)\right]=\arg\Tr Z'_\mu(x) = n \, 2\pi / 3 \,,
\end{equation}
provided
\begin{equation}
    2\cos\Lambda^- + 1 > 0 \,.
\end{equation}

For the above condition to hold we must have
\begin{equation}
    -\frac{2\pi}{3} < \Lambda^- < \frac{2\pi}{3} \,.
\end{equation}
Explicitly this implies that
\begin{equation}
    -\frac{2\pi}{3} < \omega\left[ \sigma_\mu(x) - \lambda_\mu(x) \right] < \frac{2\pi}{3} \,,
    \label{eq:expcon}
\end{equation}
where we have used equations (\ref{eq:lambda_j}) and (\ref{eq:lambda_k}). Since
\begin{align}
    \lambda_\mu(x) = n\,\frac{2\pi}{3}\,,\ \ n \in \left\{-1,0,1\right\} \,, \\
    \sigma_\mu(x) = m\,\frac{2\pi}{6 \cdot 3}\,,\ \ m \in \left[-18,18\right] \subset \Z \,,
\end{align}
we can rewrite equation~(\ref{eq:expcon}) as
\begin{equation}
    -\frac{2\pi}{3} < \omega\left[ m\,\frac{2\pi}{6 \cdot 3}  - n\,\frac{2\pi}{3}\right] < \frac{2\pi}{3} \,,
\end{equation}
which for $\omega > 0$ simplifies to
\begin{equation}
    \omega \vert  m - 6n \vert < 6 \,.
\end{equation}

Considering the extrema where $m=\pm18$ and $n=\mp1$ we require that
\begin{equation}
    \omega < \frac{1}{4} \,.
\end{equation}
In practice, we always choose small $\omega \ll \frac{1}{4}$.

\section{Centrifuge Preconditioned Smoothing}
\label{sec:centrifuge_preconditioned_smoothing}

We consider a center-vortex configuration projected from a $20^3\times40$ Luscher-Weisz $\mathcal{O}(a^2)$ mean-field-improved action pure-gauge configuration with lattice spacing $a=0.125$ fm.
This same configuration is used throughout the rest of the paper.
It is expected that the total action will increase after the vortex links have experienced centrifuge preconditioning. In general, we desire the centrifugal rotation angle $\omega$ to be small as we only wish to minimally perturb the vortex links.

\begin{figure*}[t]
    \includegraphics[width=\linewidth]{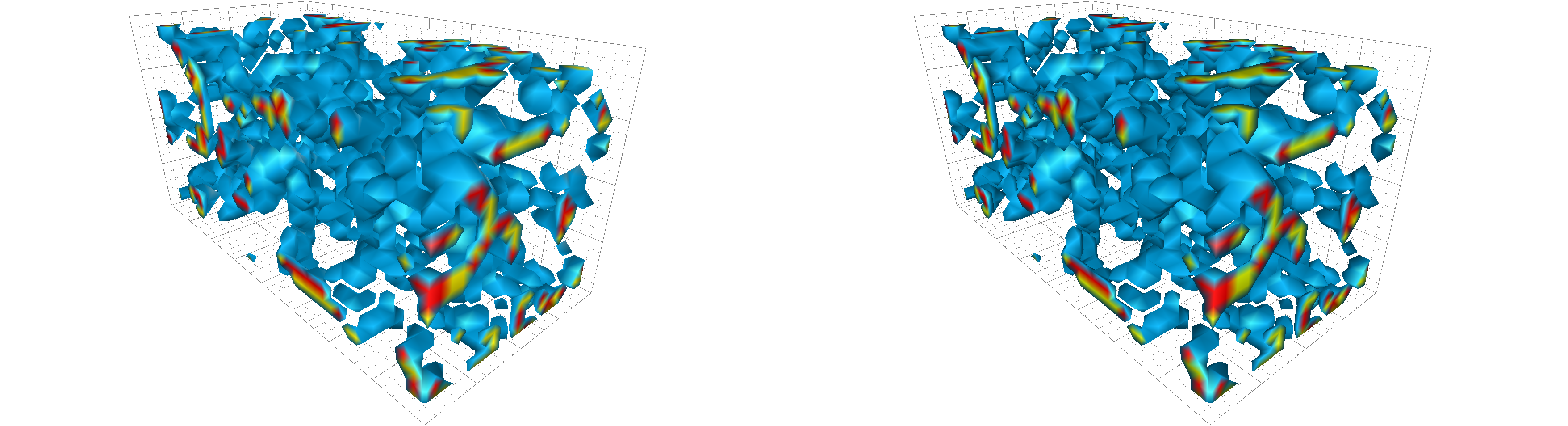}
    \caption{Action density $S(x)$ of a single time slice of a $\Zc(3)$ center-vortex gauge field before (left) and after (right) centrifuge preconditioning at $\omega=0.02$.}
    \label{S-CS}
\end{figure*}

\begin{figure*}[t]
    \includegraphics[width=\linewidth]{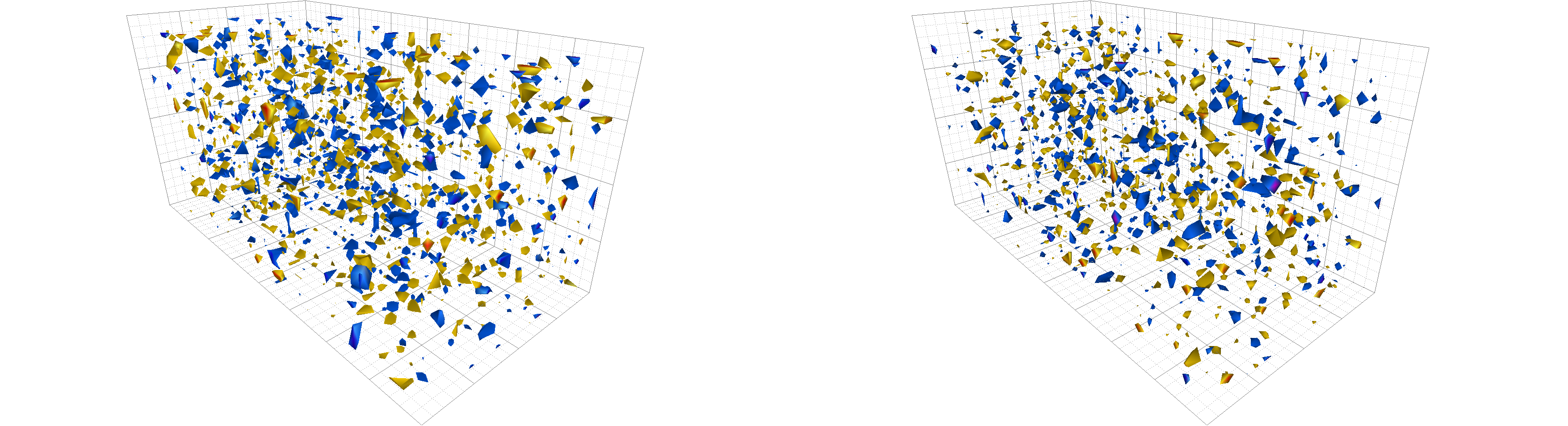}
    \caption{Topological charge density $q(x)$ of a single time slice of a $\Zc(3)$ center-vortex gauge field before (left) and after (right) centrifuge preconditioning at $\omega=0.02$.}
    \label{Q-CS}
\end{figure*}

In Figs.~\ref{S-CS} and~\ref{Q-CS} the action density $S(x)$ and topological charge density $q(x)$ of the center-vortex gauge field are compared before and after the centrifuge preconditioning at $\omega=0.02$ has been applied. This value of $\omega$ was chosen to sufficiently rotate the links away from the center whilst keeping the increase in the total action to an acceptable level.
Shown are the standard Wilson action density,
\begin{equation}
    S(x) = \frac{\beta}{2 n_c n_d (n_d - 1)}\sum_{\genfrac{}{}{0pt}{2}{\mu,\nu}{\mu\ne\nu}}\re\Tr\left[1-P_\munu(x)\right] \,,
\end{equation}
and the one-loop topological charge density,
\begin{equation}
    q(x) = \frac{1}{32\pi^2}\epsilon_{\mu\nu\rho\sigma}\Tr\left[F^{\munu}(x)F^{\rho\sigma}\right] \,,
    \label{eq:1LTopQ}
\end{equation}
where
\begin{equation}
    F_\munu(x) = \frac{1}{2ig}\left[ C_\munu(x) - C^\dagger_\munu(x)\right] \,,
    \label{eq:1LFST}
\end{equation}
and $C_\munu(x)$ is the $1\times 1$ clover term.
The action appears invariant with only a few pixels in the image changing. 
This suggests that we have successfully broken the diagonal symmetry without significantly altering the underlying center-vortex structure of the gauge field.

The topological charge density does show some more significant changes, however the physical significance of the changes in Fig.~\ref{Q-CS} are not clear.
In general, we would consider the topological charge density to have physical meaning when the gauge field is smooth enough for the Atiyah-Singer index theorem to be satisfied~\cite{Atiyah:1968mp}, such that the gluonic definition of the integrated topological charge is approximately an integer and also agrees with the fermionic definition measured by the difference of left- and right-handed zero modes of the overlap-Dirac operator~\cite{Narayanan:1997sa}. 
Previous studies show that 2 to 3 sweeps of standard stout-link smoothing at $\rho=0.1$ is required for the lattice operators to become good approximations to the physical charge~\cite{Moran:2008qd,Moran:2010rn,Teper:2022mmj}. The extremely rough nature of the projected vortex fields do not satisfy this condition. However, we do note that on a center vortex field the topological charge density necessarily correlates with the singular points of the dual vortices~\cite{Biddle:2019gke}, and that after smoothing vortex fields can generate instanton-like structures~\cite{Trewartha:2015ida}. Hence, the topological charge density on (smoothed) center vortex fields remains of interest.


\subsection{Smoothness condition}

As a measure of smoothness, we compare the mean densities of the standard Wilson action, and the reconstructed Wilson action~\cite{Bilson-Thompson:2002xlt}, given by
\begin{equation}
    \langle S \rangle  = \frac{1}{n_{\text{lat}}} \sum_x S(x) \,,
\end{equation}
and
\begin{equation}
    \langle S_\text{R} \rangle = \frac{\beta}{2 n_c n_d (n_d - 1)} \frac{1}{n_\text{lat}} \sum_{x,\mu,\nu}\Tr\left[ F_\munu(x)F_\munu(x) \right]\,,
\end{equation}
respectively.

The standard and reconstructed Wilson actions differ by $\mathcal{O}(a^6)$ terms and perturbative renormalization factors.
As the gauge field becomes smoother the perturbative contributions are suppressed and the renormalization factors tend towards 1.
Thus, the difference between the standard and reconstructed action can be used as a measure of the smoothness of the gauge field.
We consider the gauge field sufficiently smoothed when $\langle S \rangle \approx \langle S_\text{R} \rangle$.

\subsection{Smoothing in MCG}
\label{sec:smoothingMCG}

We examine centrifuge-preconditioned vortex fields that have been smoothed in MCG, starting with the Wilson flow.
The Euler method for numerically integrating the Wilson flow~\cite{Luscher:2009eq} updates links according to
\begin{equation}
    U_\mu(x,\tau) \to U_\mu(x,\tau+\epsilon) = e^{\epsilon Q_\mu(x)[U]} U_\mu(x,\tau).
\end{equation}
In effect, this is an annealed implementation of stout-link smearing~\cite{Morningstar:2003gk}, where links are updated one at a time rather than simultaneously, and the smearing parameter 
$\rho$ corresponds to the integration step size $\epsilon$.
It follows then, that flow time $\tau = n\rho$ after $n$ sweeps of smearing.
Whilst more sophisticated Runge-Kutta methods exist and have been used, we restrict our initial investigation to the Euler method.

\begin{figure}[t]
    \includegraphics[width=\linewidth]{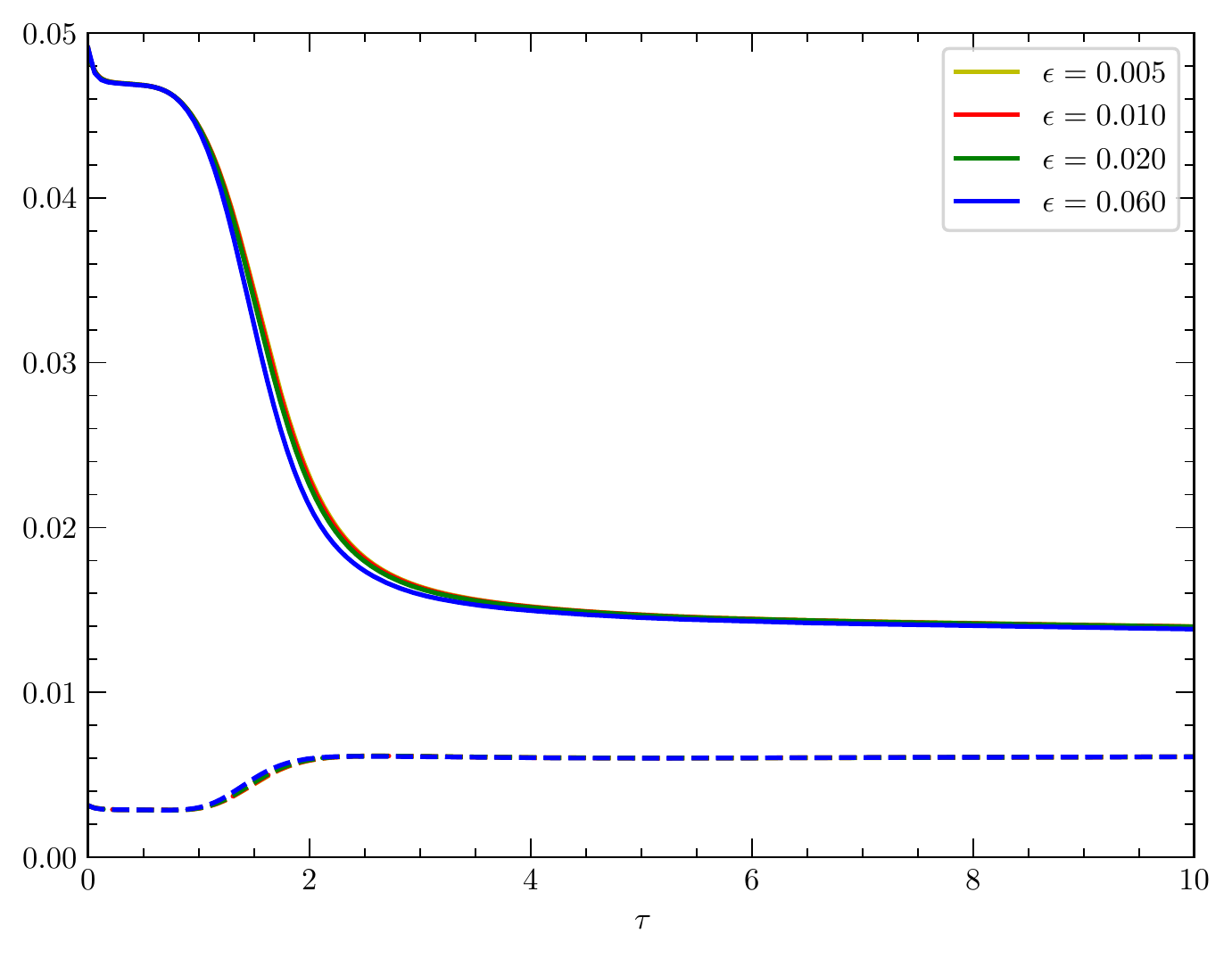}
    \caption{$\langle S \rangle$ (solid) and $\langle S_\text{R} \rangle$ (dashed) as a function of Wilson flow time $\tau$ for integration step size $\epsilon=0.005,\,0.01,\,0.02,\,0.06$.}
    \label{fig:wf}
\end{figure}

Fig.~\ref{fig:wf} shows the mean densities $\langle S \rangle$ and $\langle S_\text{R} \rangle$ of the centrifugal preconditioned gauge field as a function of Wilson flow time $\tau$ computed with Euler integration step sizes $\epsilon=0.06,\,0.02,\,0.01,\,0.005$.
The flow is no longer invariant and smooths the gauge field, however the direct smearing of centrifuge preconditioned vortex fields is insufficient to bring $\langle S \rangle$ and $\langle S_\text{R} \rangle$ into agreement.
The field remains rough and does not satisfy the smoothness condition above, required for the overlap-Dirac operator to be well-defined.

In the spirit of gradient flow, we now turn to annealed $U$-link smearing (AUS) with small $\alpha=0.02$.
As mentioned previously, AUS is identical in form to APE smearing, but the links are effectively updated one at a time rather than simultaneously.
We employ the over-improvement formalism where the staples term is given in equation (\ref{eq:over-improvement}).
We choose $\epsilon=-0.25$ as per Ref.~\cite{Moran:2008ra}. We can consider AUS coupled with either unit circle projection (AUS+UCP) or MaxReTr reuniterization (AUS+MaxReTr).

\begin{figure}[t]
    \subfloat[Unit circle projection.\label{fig:CS0.02-AUSUCP0.02}]{\includegraphics[width=0.9\columnwidth]{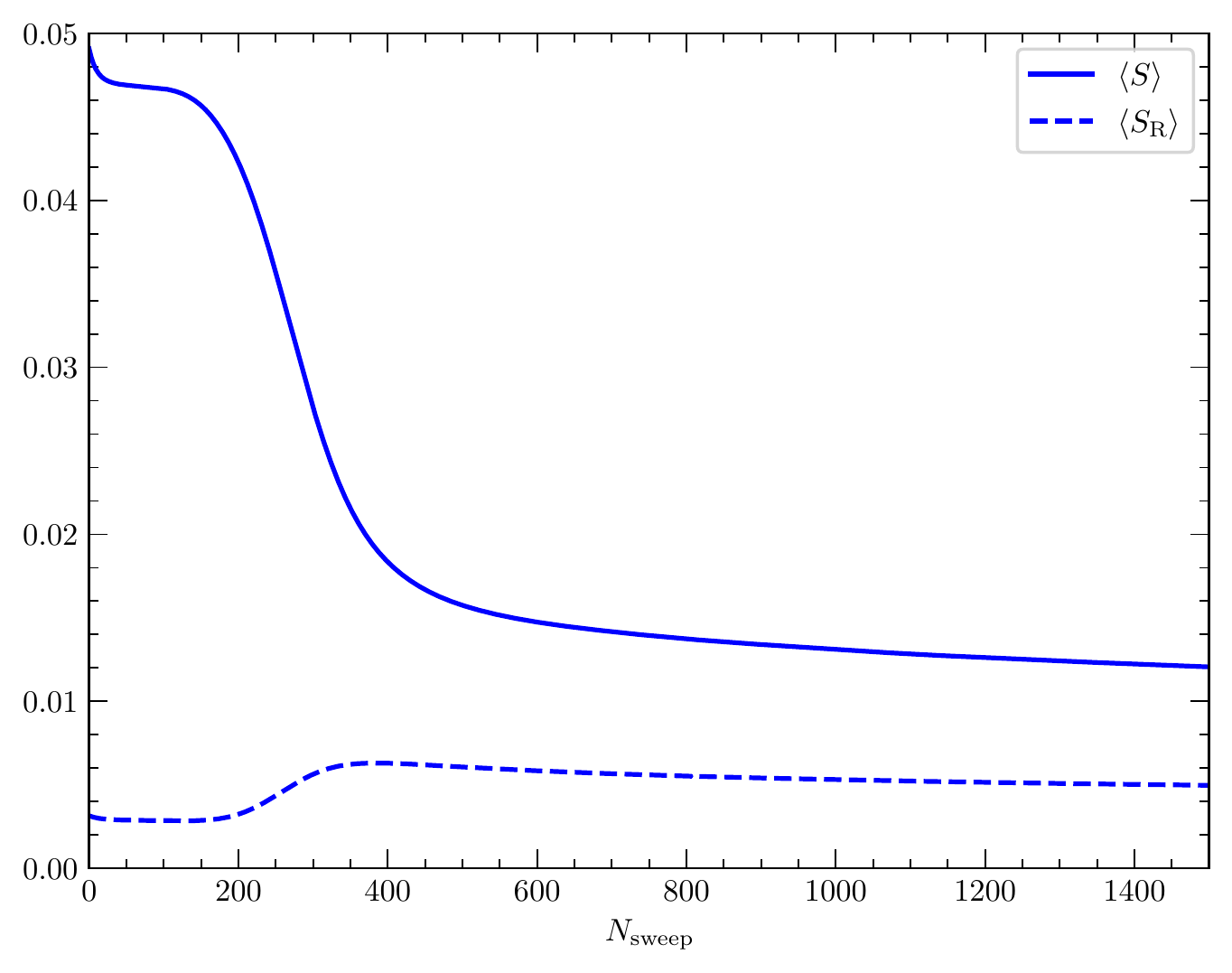}}
    
    \subfloat[MaxReTr reuniterisation.\label{fig:CS0.02-AUSMRT0.02}]{\includegraphics[width=0.9\columnwidth]{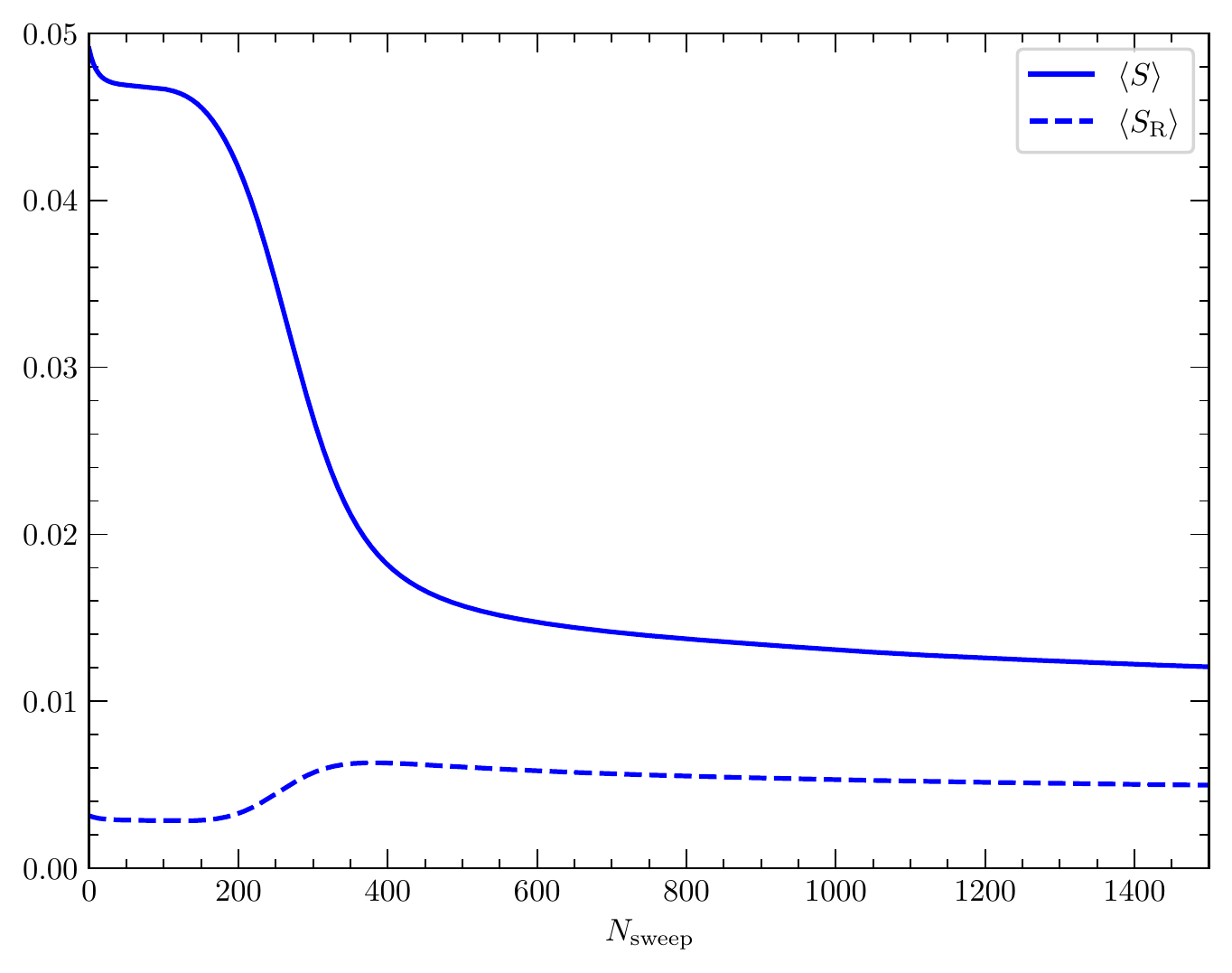}}
    
    \caption{$\langle S \rangle$ and $\langle S_\text{R} \rangle$ as a function of $N_\text{sweep}$ iterations of over-improved AUS at $\epsilon=-0.25$ and $\omega=0.02$ using (a) unit circle projection and (b) MaxReTr reuniterization, applied to a centrifuge preconditioned gauge field.}
    \label{fig:CS0.02-AUS0.02}
\end{figure}

Fig.~\ref{fig:CS0.02-AUSUCP0.02} shows the results for AUS with unit circle projection. Again (as with the Wilson flow above) on centrifugal preconditioned vortex fields the gauge field is smoothed, but insufficiently to bring $\langle S \rangle$ and $\langle S_\text{R} \rangle$ into agreement and satisfy the required smoothness condition. 
We find similar results in Fig.~\ref{fig:CS0.02-AUSMRT0.02} where the MaxReTr reuniterization has been used instead.

In all three of the cases above the smeared links remain diagonal. This is in some sense expected, as taking a linear combination of the diagonal matrices will result in a diagonal matrix for the staples, such that the smoothed link will also remain within the diagonal subgroup of $\SU(3).$ This means that the smoothing process is unable to form links that encompass the full manifold of the special unitary group.

The inability of these algorithms to smear the diagonal elements of a particular link into its off-diagonal elements appears to present a fundamental limitation to the amount of smoothing which can be achieved.
As such, it seems necessary to employ an algorithm which is able to mix the diagonal and non-diagonal elements of a link.
To this end, rather than starting from maximal center gauge we consider the addition of a random gauge transformation.

\subsection{Smoothing in random gauge}
\label{sec:smoothingRandom}

\begin{figure}[t]
    \includegraphics[width=0.9\columnwidth]{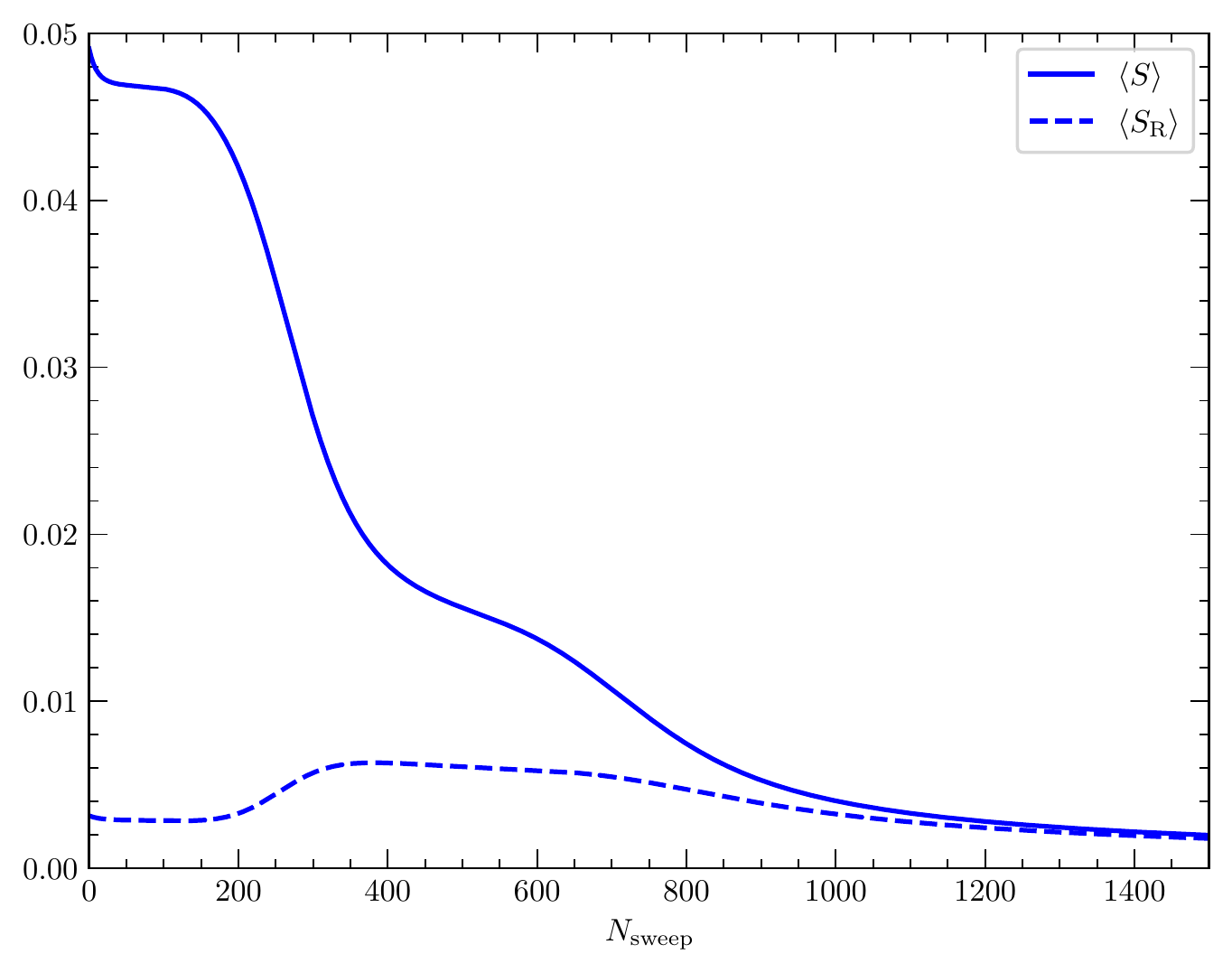}
    \caption{$\langle S \rangle$ and $\langle S_\text{R} \rangle$ as a function of $N_\text{sweep}$ iterations of over-improved AUS at $\epsilon=-0.25$ and $\omega=0.02$ using the MaxReTr reuniterization applied to a centrifuge preconditioned gauge field which has been transformed to a random gauge.}
    \label{fig:CS0.02-R-AUSMRT0.02}
\end{figure}

The Wilson flow and AUS with unit circle projection are gauge equivariant, which is to say for some smoothing process $\mathcal{S}$ and gauge transformation
\begin{equation}
    U_\mu(x) \to U^G_\mu(x) = G(x)U_\mu(x)G^\dagger(x+\muhat)
\end{equation}
that
\begin{equation}
    \mathcal{S}\left\{U^G_\mu(x)\right\} = G(x) \mathcal{S} \left\{ U_\mu(x) \right\} G^\dagger(x+\muhat)\,.
\end{equation}
As the linear combination of two diagonal matrices remains diagonal, this gauge equivariance prevents the analytic smoothing algorithms from leaving the diagonal subgroup of $\SU(3).$
This is not the case in general for AUS with MaxReTr reuniterization.
As such, we repeat the AUS+MaxReTr calculation with identical parameters, but this time we have transformed the centrifuge preconditioned gauge field to a random gauge before smoothing.
We see in Fig.~\ref{fig:CS0.02-R-AUSMRT0.02} that now the gauge field can be sufficiently smoothed, achieving agreement between the action and reconstructed action with enough ($N_\text{sweeps} > 1000$) sweeps of smoothing.
We find that this smoothness condition is sufficient for the overlap-Dirac operator to be well-defined.

\section{Vortex-preserved annealing}
\label{sec:vpas}

One of the stated goals of finding a method to smooth center vortex gauge fields was to preserve the underlying vortex structure.
To this end, we introduce \emph{vortex-preserved annealed smoothing} (VPAS) via an additional accept/reject step, which, in principle, can be applied to any iterative smoothing algorithm.

Let us first consider VPAS applied to center vortices in MCG.
The AUS algorithm is run as usual to produce a candidate link in $\SU(3),$
\begin{equation}
    Z'_\mu(x) = \mathcal{P}_{\SU(3)}\left\{V^{(n)}_\mu(x)\right\}\,.
\end{equation}
The updated link is then given by
\begin{equation}
    Z^{(n+1)}_\mu(x)=
    \begin{cases}
        Z'_\mu(x) \text{ if } \mathcal{P}_{\Zc(3)}\left\{Z'_\mu(x)\right\} = Z_\mu(x) \,,\\
        Z^{(n)}_\mu(x) \text{ otherwise,}
    \end{cases}
\end{equation} 
which is to say a candidate link is only accepted if it projects back to the original center vortex link using Eq.~(\ref{eq:centre_project}).
In the case where the original center vortex link as undergone an arbitrary gauge transformation
\begin{equation}
    Z_\mu(x) \to G(x)Z_\mu(x)G^\dagger(x+\muhat)
\end{equation}
the acceptance condition becomes
\begin{equation}
    \mathcal{P}_{\Zc(3)}\left\{ G^\dagger(x) Z'_\mu(x) G(x+\muhat)\right\} = Z_\mu(x)\,,
\end{equation}
where the inverse of the original gauge transformation is applied to the candidate link. Note that in either case, the $\Zc(3)$ projection test is performed directly without reasserting the MCG condition in Eq.~(\ref{eq:MCG}).

We now consider how VPAS with MaxReTr reuniterization applies to $\Zc(3)$ center vortex gauge field configurations by studying the outcome of the first sweep.
We first examine the MCG case, where no centrifuge preconditioning has been applied.
We study three quantities in our analysis:
\begin{itemize}
\item $p_\text{in}$ ($n_\text{in}$), the proportion (absolute number) of links which satisfy $\vert \phi \vert > \frac{\pi}{2}$ (required to perturb the vortex link),
\item $p_\text{diff}$ ($n_\text{diff}$), the proportion (absolute number) of links for which $Z^{(1)}_\mu(x) \ne Z_\mu(x)$ (accounting for the possibility that although the $\phi$ condition is satisfied, it is still possible the projected link could be the same as the original), and
\item $p_\text{pass}$ ($n_\text{pass}$), the proportion (absolute number) of candidate links which pass the vortex preservation step, given $Z^{(1)}_\mu(x) \ne Z_\mu(x)$.
\end{itemize}
These definitions necessitate the condition $p_\text{in} \geqslant p_\text{diff} \geqslant p_\text{pass}$ holds.\footnote{See Section~\ref{sec:note_on_cooling} for an explanation of the apparent violation of this condition at $\alpha=1.0$ in Table~\ref{t:VO.vpa}.}

\begin{figure}[t]
    \includegraphics[width=0.9\columnwidth]{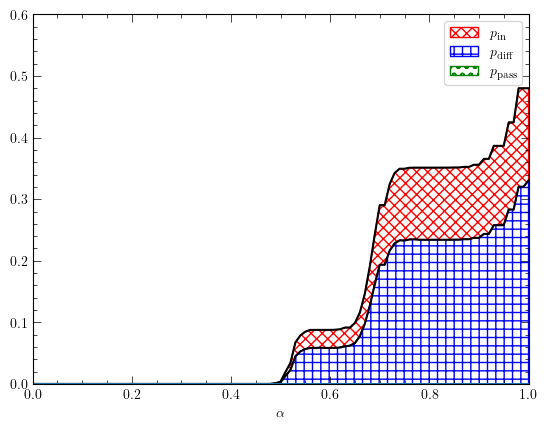}
    \caption{The proportion of all possible combinations of links, weighted by multiplicity, which satisfy $\vert \phi \vert > \frac{\pi}{2}$ ($p_\text{in}$), $Z^{(1)}_\mu(x) \ne Z_\mu(x)$ ($p_\text{diff}$), and pass the vortex preservation step given $Z^{(1)}_\mu(x) \ne Z_\mu(x)$ ($p_\text{pass}$). Note $p_\text{pass}=0$ for all $\alpha$}
    \label{fig:all_combo_vpas}
\end{figure}

In Fig.~\ref{fig:all_combo_vpas}, we compute these values for all combination of links, once again in the $\beta \to 0$ limit where a given link as an equal probability to be one of the three center phases, and weight each combination by its multiplicity.
Most strikingly, $p_\text{pass}=0$ for all values of $\alpha$.
This implies that if the updated link is different from the original, the phase of its trace will always fall outside the sector which center projects to the original link.

\begin{figure}[t]
    \includegraphics[width=0.9\columnwidth]{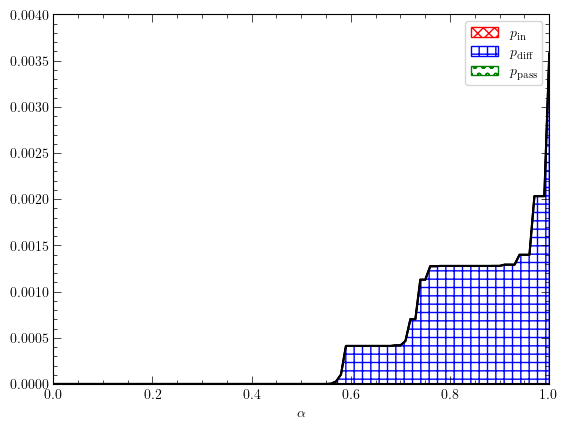}
    \caption{The proportion of links of a $\Zc(3)$ pure gauge field configuration, which satisfy $\vert \phi \vert > \frac{\pi}{2}$ ($p_\text{in}$), $Z^{(1)}_\mu(x) \ne Z_\mu(x)$ ($p_\text{diff}$), and pass the vortex preservation step given $Z^{(1)}_\mu(x) \ne Z_\mu(x)$ ($p_\text{pass}$). Note $p_\text{pass} = 0$ for all $\alpha$ and $p_\text{in} = p_\text{diff}$ for all $\alpha < 1$.}
    \label{fig:VO.vpa}
\end{figure}

We repeat this analysis on a true $\Zc(3)$ projected gauge field configuration. 
The proportions are presented in Fig.~\ref{fig:VO.vpa}, whilst the absolute values are tabulated at intervals of 0.1 for $\alpha$ in Table~\ref{t:VO.vpa}.
Consistent with the analysis presented in Fig~\ref{fig:all_combo_vpas}, we see $p_\text{pass} = 0$ for all values of $\alpha$ and similar shaped curves for $p_\text{diff}(\alpha)$.
However, unlike the previous analysis, we have that $p_\text{in} = p_\text{diff}$ for all $\alpha$.

\begin{figure}[t]
    \includegraphics[width=0.9\columnwidth]{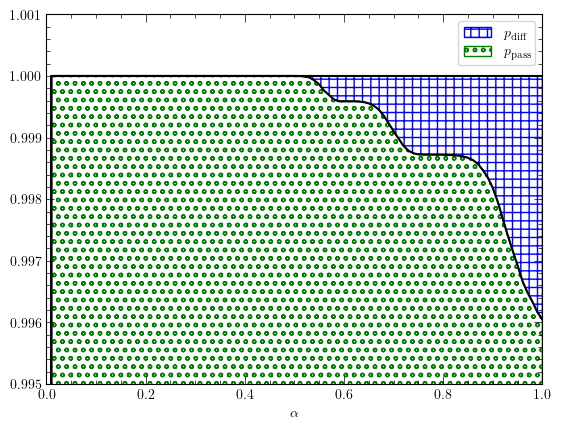}
    \caption{The proportion of links of a centrifuge preconditioned $\Zc(3)$ pure gauge field configuration, which satisfy $Z^{(1)}_\mu(x) \ne Z_\mu(x)$ ($p_\text{diff}$), and pass the vortex preservation step given $Z^{(1)}_\mu(x) \ne Z_\mu(x)$ ($p_\text{pass}$). Note that $p_\text{in} = 1$ for all $\alpha$.}
    \label{fig:VO-CS0.02.vpa}
\end{figure}

The same analysis is performed after centrifuge preconditioning and presented in Fig.~\ref{fig:VO-CS0.02.vpa} and Table~\ref{t:VO-CS0.02.vpa}, where $p_\text{in}$ has been dropped as this condition only applies to an unconditioned $\Zc(3)$ gauge field where all links are proportional to the identity. 
Here (aside from the trivial $\alpha=0$ case), we see not only that every updated link is different from the original at all values of $\alpha$, but also that every candidate link passes the vortex preservation step below $\alpha\approx0.5$ and almost all ($>99.5\%$) at larger values of $\alpha$.
While this suggest that a vortex-preservation step is not required for $\alpha < 0.5$, we note this is only for the first sweep following centrifugal preconditioning.
Eventually, the vortex-preservation step does have an effect on the smoothing process.

Performing the same one-sweep analyses after a random gauge transform has been applied produces near-identical results, and for the sake of brevity will not be presented herein.
This is not unexpected, as we are only considering the first sweep.
As we can see comparing Figures~\ref{fig:CS0.02-AUS0.02} and~\ref{fig:CS0.02-R-AUSMRT0.02} the random gauge transformation only begins to have significance after several sweeps.

\begin{figure}[t]
    \subfloat[No random gauge transformation.]{\includegraphics[width=0.9\columnwidth]{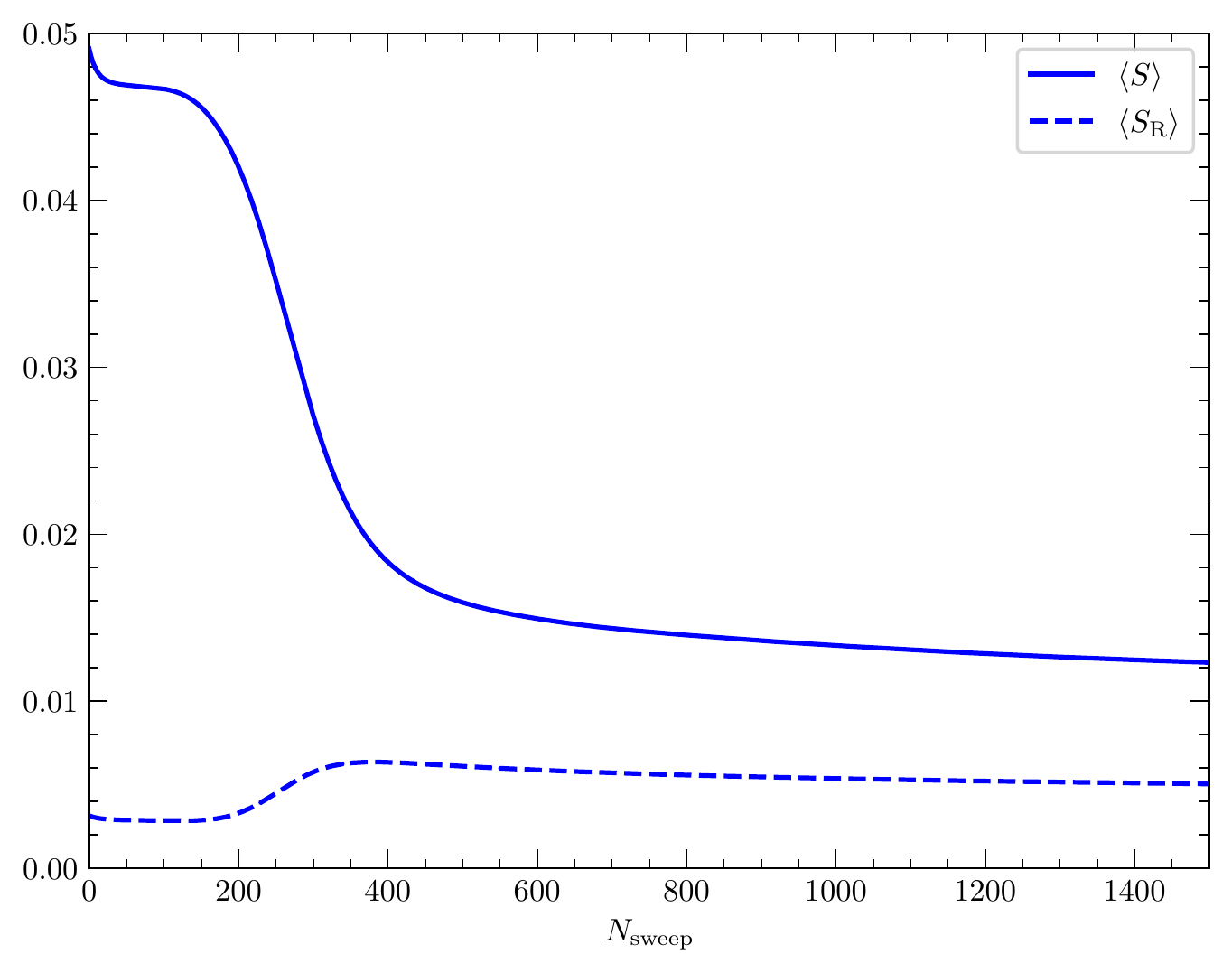}}
    \\
    \subfloat[Random gauge transformation.]{\includegraphics[width=0.9\columnwidth]{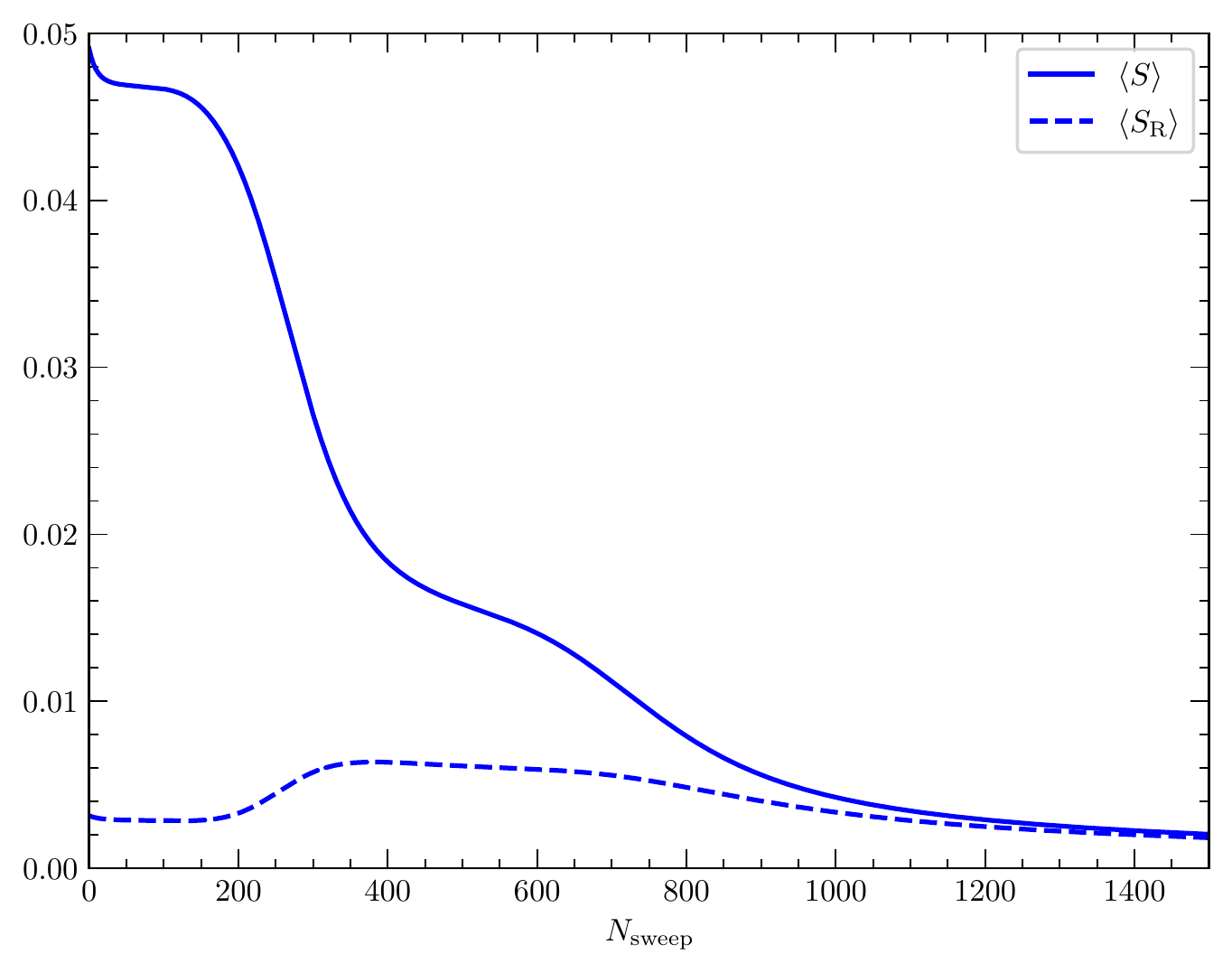}}
    \caption{$\langle S \rangle$ and $\langle S_\text{R} \rangle$ as a function of $N_\text{sweep}$ iterations of over-improved VPAS at $\epsilon=-0.25$ and $\omega=0.02$ using the MaxReTr reuniterization applied to a centrifuge preconditioned gauge field (a) without and (b) with a random gauge transformation applied after preconditioning and before smoothing.}
    \label{fig:S-VO-CS0.02-AUSVPMRT0.02_RvMCG}
\end{figure}

Applying VPAS with MaxReTr reuniterization to a centrifuge preconditioned gauge field, we find similar results, presented in Fig.~\ref{fig:S-VO-CS0.02-AUSVPMRT0.02_RvMCG}, to what we have seen with regular AUS.
Once again, applying a random gauge transform to the field is necessary to achieve sufficient smoothing.

\begin{ruledtabular}
    \begin{table}
        \caption{The number of links of a $\Zc(3)$ pure gauge field configuration satisfying various conditions. $n_\text{in}$ counts the links which satisfy $\vert \phi \vert > \frac{\pi}{2}$. $n_\text{diff}$ counts the links for which $Z^{(1)}_\mu(x) \ne Z_\mu(x)$. $n_\text{pass}$ counts the links satisfying the preservation condition given $Z^{(1)}_\mu(x) \ne Z_\mu(x)$. The $20^3 \times 40$ lattice has 1,280,000 links.}
        \label{t:VO.vpa}
        \begin{tabular}{ c c c c }
            \noalign{\smallskip}
            $\alpha$ & $n_\text{in}$ & $n_\text{diff}$ & $n_\text{pass}$\\
            \noalign{\smallskip}
            \hline
            \noalign{\smallskip}
            0.4 &  \hfill    0 & \hfill    0 & \hfill 0\\
            0.5 &  \hfill    0 & \hfill    0 & \hfill 0\\
            0.6 &  \hfill  527 & \hfill  527 & \hfill 0\\
            0.7 &  \hfill  535 & \hfill  535 & \hfill 0\\
            0.8 &  \hfill 1635 & \hfill 1635 & \hfill 0\\
            0.9 &  \hfill 1636 & \hfill 1636 & \hfill 0\\
            1.0 &  \hfill 4570 & \hfill 4582 & \hfill 0\\
        \end{tabular}
    \end{table}
\end{ruledtabular}

\begin{ruledtabular}
    \begin{table}
        \caption{The number of links of a centrifuge preconditioned $\Zc(3)$ pure gauge field configuration, which satisfy $\vert \phi \vert > \frac{\pi}{2}$ ($p_\text{in}$), $Z^{(1)}_\mu(x) \ne Z_\mu(x)$ ($p_\text{diff}$), and pass the vortex preservation step given $Z^{(1)}_\mu(x) \ne Z_\mu(x)$ ($p_\text{pass}$). The lattice has 1280000 links.}
        \label{t:VO-CS0.02.vpa}
        \begin{tabular}{ c c c }
            \noalign{\smallskip}
            $\alpha$ & $n_\text{diff}$ & $n_\text{pass}$\\
            \noalign{\smallskip}
            \hline
            \noalign{\smallskip}
            0.1 & 1280000 & 1280000\\
            0.2 & 1280000 & 1280000\\
            0.3 & 1280000 & 1280000\\
            0.4 & 1280000 & 1280000\\
            0.5 & 1280000 & 1280000\\
            0.6 & 1280000 & 1279473\\
            0.7 & 1280000 & 1278853\\
            0.8 & 1280000 & 1278363\\
            0.9 & 1280000 & 1277673\\
            1.0 & 1280000 & 1274937\\
        \end{tabular}
    \end{table}
\end{ruledtabular}

\section{Smoothing method comparison}
\label{sec:comparison}

Throughout the previous sections, we have arrived at three viable smoothing methods for $\Zc(3)$ center vortex gauge fields.
In Section~\ref{sec:ape_analytic} we showed that an APE-style smearing algorithm can only alter a $\Zc(3)$ vortex field provided the smearing parameter $\alpha$ is sufficiently large. Furthermore, in Section~\ref{sec:maxretr_proj} we found that the degree of smoothing was only sufficient if the vortex field had undergone a random gauge transformation and the MaxReTr reuniterization was employed. Choosing also to employ the over-improvement formalism at $\epsilon=-0.25$ with smearing parameter $\alpha=0.7$ and, implementing the algorithm in an annealed manner, we have arrived at our first smoothing recipe which we denote throughout this section as `AS' for annealed smoothing.

In the spirit of approaching the gradient flow, we showed in Section~\ref{sec:centrifuge_preconditioned_smoothing} that the use of a small smearing parameter $\alpha < \alpha_\text{min}$ is enabled by centrifuge preconditioning the vortex field. A random gauge transformation is necessarily applied after preconditioning to achieve the required level of smoothing to define our second recipe which we denoted `CP' for centrifuge preconditioning. We choose an AUS smearing parameter of $\alpha=0.02$ applied to a random gauge transformed, centrifuge preconditioned gauge field with rotation angle $\omega=0.02.$ Finally, in our third recipe, denoted `VP' for vortex preservation, we include the vortex preservation step in what is otherwise identical to our second recipe.

In addition to the resultant gauge fields of each smoothing recipe, we also consider the original $\Zc(3)$ gauge field (denoted `VO' for vortex-only). In summary, we have four gauge fields to compare,
\begin{description}
\item[VO] original vortex-projected gauge field, 
\item[AS] large $\alpha$, APE-style, random-gauge-transformed annealed smoothing,
\item[CP] as for AS except with small $\alpha$ and centrifuge preconditioning,
\item[VP] as for CP but with vortex preservation step applied.
\end{description}

We choose the number of AUS sweeps (20 for AS, and 1190 for CP and V), such that the different gauge fields have approximately matched total actions.
See Table~\ref{t:recipes} for a summary of each algorithm.

\begin{figure*}[t]
    \includegraphics[width=\linewidth]{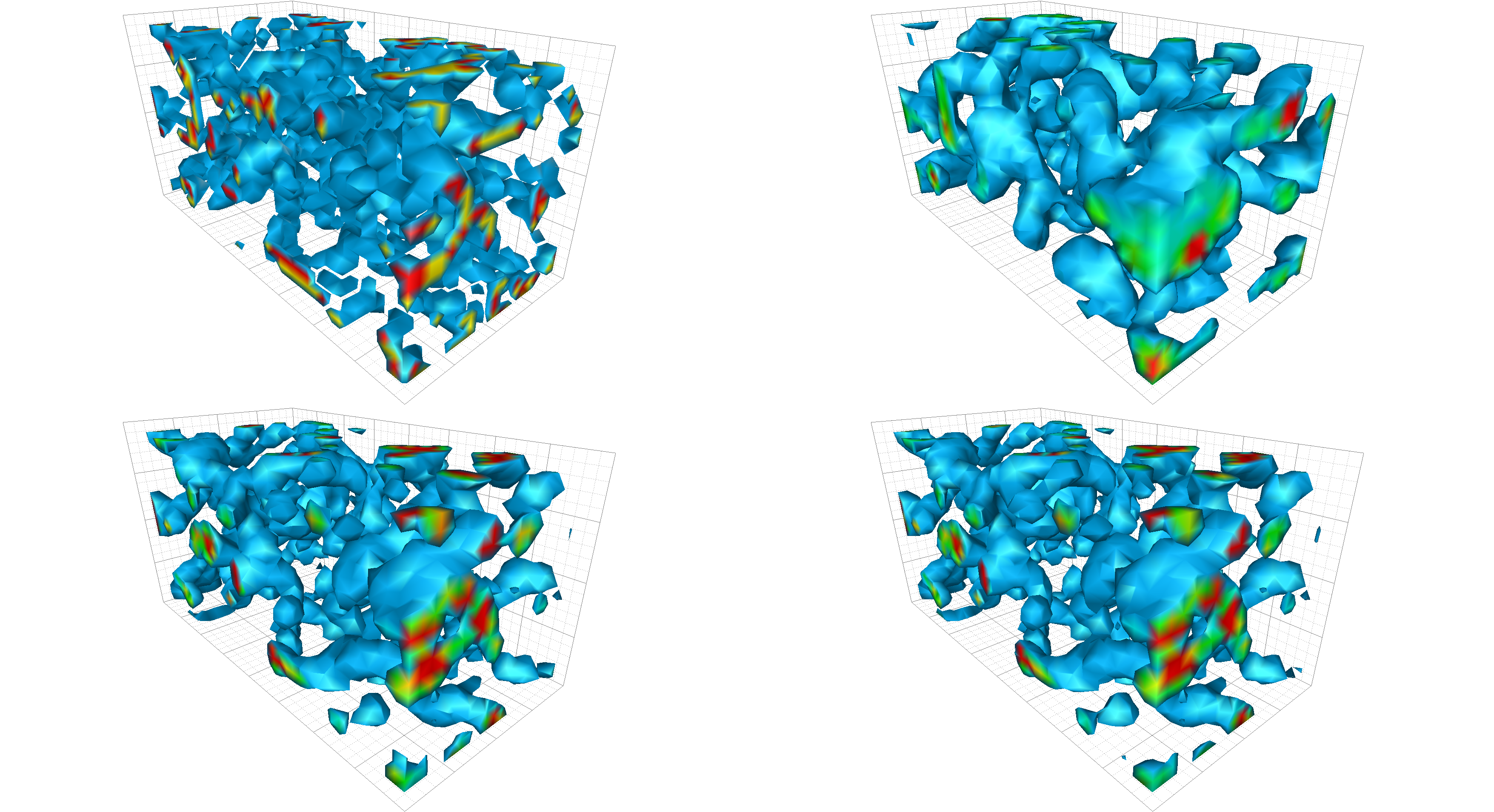}
    \caption{Action density of a single time slice after smoothing. Clockwise from top left: VO, AS, CP, VP.}
    \label{S-Density}
\end{figure*}

We present visualizations of the respective action densities in Fig.~\ref{S-Density} and compute their correlations where $C^S_{XY}$ given by
\begin{equation}
  C^S_{XY} = \frac{\langle S_X(x) \, S_Y(x) \rangle}{\sqrt{\langle S^2_X(x) \rangle} \, \sqrt{\langle S^2_Y(x) \rangle}}\,,
  \label{eq:correlation_s}
\end{equation}
is the correlation between $S_X(x)$ and $S_Y(x)$ for respective smearing processes $X$ and $Y$.
These are presented in Table~\ref{t:correlation_s}.

\begin{figure*}[t]
    \includegraphics[width=\linewidth]{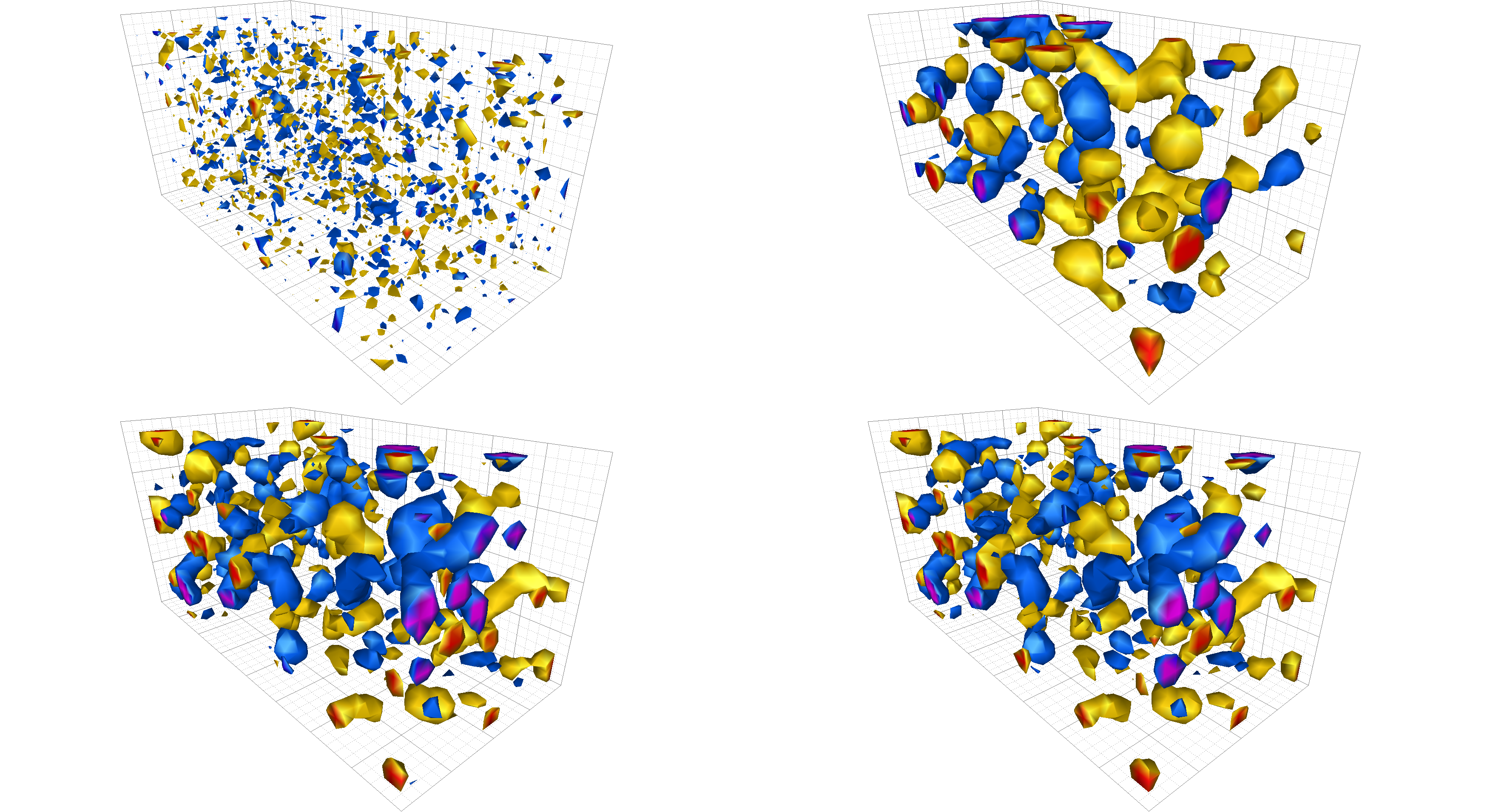}
    \caption{Topological charge density of a single time slice after smoothing. Clockwise from top left: VO, AS, CP, VP.}
    \label{Q-Density}
\end{figure*}

Similarly, we present visualizations of the respective topological charge densities in Fig.~\ref{Q-Density} and compute their correlations where $C^q_{XY}$ given by
\begin{equation}
    C^q_{XY} = \frac{\langle q_X(x) \, q_Y(x) \rangle}{\sqrt{\langle q^2_X(x) \rangle} \, \sqrt{\langle q^2_Y(x) \rangle}}\,,
\end{equation}
is the correlation between $q_X(x)$ and $q_Y(x)$ for respective smearing processes $X$ and $Y$.
These are presented in Table~\ref{t:correlation_q}.

The superior similarity of the CP and VP action densities to the original vortex field evident in the visualizations, as compared to AS, indicates not only that the use of a small smearing parameter is desirable, but it is fundamentally important in preserving the underlying vortex structure.
With regard to the action density, the numerical correlation of VP with the original vortex field is slightly higher (by $\sim 2\%$) as compared to AS and CP.

Visually, the topological charge density for VO is qualitatively different to the three smoothed fields in terms of the size and number of objects. The numerical comparison of the topological charge densities indicates they are essentially uncorrelated, with the exception of CP and VP which do show a strong positive correlation.

\begin{ruledtabular}
    \begin{table}
        \caption{Summary of smoothing recipes. Steps are applied from left to right starting with the $\Zc(3)$ center-vortex configuration in MCG. C indicates centrifuge preconditioning with rotation angle $\omega$. R indicates the application of a random gauge transformation. $N_\text{AUS}$ indicates the number of sweeps of AUS at smearing parameter $\alpha$. V indicates if a vortex-preservation step was included in the AUS smearing.}
        \label{t:recipes}
        \begin{tabular}{ l c c c c c c }
            \noalign{\smallskip}
            Algorithm & C & $\omega$ & R & $N_\text{AUS}$ & $\alpha$ & V \\
            \noalign{\smallskip}
            \hline
            \noalign{\smallskip}
            VO & $\times$ & - & $\times$ & 0 & - & - \\
            AS & $\times$ & - & $\checkmark$ & 20 & 0.7 & $\times$ \\
            CP & $\checkmark$ & 0.02 & $\checkmark$ & 1190 & 0.02 & $\times$ \\
            VP & $\checkmark$ & 0.02 & $\checkmark$ & 1190 & 0.02 & $\checkmark$ \\
        \end{tabular}
    \end{table}
\end{ruledtabular}

\begin{ruledtabular}
    \begin{table}
        \caption{The correlation $C^S_{XY}$ of action densities $S_X(x)$ and $S_Y(x)$ of the gauge fields after respective smoothing algorithms $X$ and $Y$ have been applied. }
        \label{t:correlation_s}
        \begin{tabular}{ l c c c c}
            \noalign{\smallskip}
                & VO     & AS       & CP    & VP       \\
            \noalign{\smallskip}
            \hline
            \noalign{\smallskip}
            VO  & 1.000 & \hfill 0.397 & \hfill 0.396 & \hfill 0.403 \\
            AS  & -     & \hfill 1.000 & \hfill 0.519 & \hfill 0.513 \\
            CP  & -     & -            & \hfill 1.000 & \hfill 0.942 \\
            VP  & -     & -            & -            & \hfill 1.000 \\
        \end{tabular}
    \end{table}
\end{ruledtabular}

\begin{ruledtabular}
    \begin{table}
        \caption{The correlation $C^q_{XY}$ of topological charge densities $q_X(x)$ and $q_Y(x)$ of the gauge fields after respective smoothing algorithms $X$ and $Y$ have been applied. }
        \label{t:correlation_q}
        \begin{tabular}{ l c c c c}
            \noalign{\smallskip}
                & VO     & AS       & CP    & VP       \\
            \noalign{\smallskip}
            \hline
            \noalign{\smallskip}
            VO  & 1.000 & \hfill 0.010 & \hfill -0.001 & \hfill 0.002 \\
            AS  & -     & \hfill 1.000 & \hfill 0.009 & \hfill 0.001 \\
            CP  & -     & -            & \hfill 1.000 & \hfill 0.884 \\
            VP  & -     & -            & -            & \hfill 1.000 \\
        \end{tabular}
    \end{table}
\end{ruledtabular}
        

\section{Summary}
\label{sec:sum}

Throughout this work we have studied the application of a variety of $\SU(3)$ gauge field smoothing methods to $\Zc(3)$ center vortex gauge fields, with an aim to achieve sufficient smoothness so as to be able to meaningfully evaluate the overlap-Dirac operator. An additional aim is to preserve (as much as possible) the original vortex structure identified in MCG.
Due to the proportionality of the vortex-field links to the identity, a naive application of traditional smoothing algorithms is either ineffectual or limited, containing subtle issues which are not obviously manifest.

To overcome these issues, we introduced a novel method, centrifuge preconditioning, which perturbs the center elements from $\Zc(3)$ into the $\U(1)\times \U(1)\times \Zc(3)$ diagonal subgroup of $\SU(3).$ The centrifuge preconditioning step is constructed in a manner that breaks the proportionality of the links to the identity while preserving the original vortex information.

Agreement between the action and reconstructed action is set as the condition for sufficient smoothness in order to employ overlap fermions on the smoothed vortex field. 
The amount of smoothing that can be obtained with analytic methods is fundamentally limited by the gauge equivariant property of these methods, which even with centrifuge preconditioning remain within the diagonal subgroup of $\SU(3)$ (up to a gauge transformation).

It is only through the application of a random gauge transform together with the MaxReTr reuniterization -- an update-based method that is not gauge equivariant -- as part of an APE-based annealed smoothing formalism, that it becomes possible to depart from the diagonal subgroup and expand the smoothed links to the greater part of the $\SU(3)$ group manifold.

Additional to centrifuge preconditioning, to preserve the vortex structure throughout the annealed smoothing process, the concept of a vortex preservation step was introduced.
This consists of an accept/reject step within the annealed smoothing process, where the update for a given link is only accepted if the argument of the trace projects to the same $\Zc(3)$ element as the center phase identified in the MCG of the original gauge field.

Based on the above, three smoothing recipes were formulated (AS, CP, VP) which along with the pure vortex field (VO) were compared (refer to Table~\ref{t:recipes} for a summary). With regard to the action density, the visualizations in Fig.~\ref{S-Density} show the CP and VP algorithms produce action densities resembling the original vortex structure. Considering the quantitative measure of Eq.~(\ref{eq:correlation_s}), all three smoothing recipes were found to have a similar correlation with the original field, with VP having the highest of the three by a small margin. On such rough fields, the microscopic structure of the topological charge density appears to be volatile and as a result there is essentially no correlation between three of the four fields examined. The exceptional pair is CP and VP, which produce highly correlated topological charge densities, the only difference between these two recipes being that VP includes the vortex-preserved annealing step ensuring that the argument of the trace of the links projects to the original center element. 

The conclusion is that the centrifuge preconditioning and vortex-preserved annealing techniques enable the successful smoothing of vortex fields, and will be studied further in future work.

\begin{acknowledgments}

We thank the PACS-CS Collaboration for making their configurations available via the International Lattice Data Grid (ILDG).
This research was undertaken with resources provided by the Pawsey Supercomputing Centre through the National Computational Merit Allocation Scheme with funding from the Australian Government and the Government of Western Australia. Additional resources were provided from the National Computational Infrastructure (NCI) supported by the Australian Government through Grant No. LE190100021 via the University of Adelaide Partner Share.
This research is supported by Australian Research Council through Grants No. DP190102215 and DP210103706.
WK is supported by the Pawsey Supercomputing Centre through the Pawsey Centre for Extreme Scale Readiness (PaCER) program.

\end{acknowledgments}

\appendix

\section{Generators of $\SU(3)$}
\label{sec:generatorsSU3}

The eight traceless anti-Hermitian matrices $T^a$ are the generators of $\SU(3),$ and are proportional to the Gell-Mann matrices.
We choose the normalization condition $\Tr \left[T^a\,T^b\right]=-\frac{1}{2}\delta^{ab}$ to ensure the structure constants $f^{abc}$ defined by  $\left[T^a,T^b\right]=f^{abc}\,T^c$ are real and totally antisymmetric in the indices.

Up to a center phase factor, the diagonal subgroup of $\SU(3)$ is spanned by the subset of generators $\{ T^3, T^8 \},$ where $T^3$ and $T^8$ are diagonal. 
We can write an element of the diagonal subgroup as
\begin{equation}
\exp\left(n\frac{2\pi i}{3}\right)\exp\left( a_3T^3 + a_8T^8 \right), 
\end{equation}
where $n \in \{ -1,0,1 \},$ and $a_3,a_8 \in \R.$ The subgroup is Abelian as $[T^3,T^8] = 0,$ and it is straightforward to see that it is isomorphic to $\U(1)\times \U(1) \times \Zc(3).$

\section{Derivation of $\alpha_\text{min}$ for MaxReTr reuniterization within the over-improvement formalism}
\label{sec:alpha_min_derivation}

Without loss of generality, let $Z_\mu(x) = \I$. Hence, $L^1$ from equation (\ref{eq:L_gauge_invariant}) becomes
\begin{equation}
    L^1=(1-\alpha) + \frac{\alpha}{6}\Sigma^\dagger_\mu(x)\,,
\end{equation}
which explicitly within the over-improvement formalism is
\begin{equation}
    L^1=(1-\alpha) + \frac{\alpha}{6}\left\{\left(\frac{5-2\epsilon}{3}\right)S_\mu(x) + \left(\frac{\epsilon-1}{12u^2_0}\right)R_\mu(x)\right\}\,,
    \label{eq:L_overimp_explicit}
\end{equation}
where $S_\mu(x)$ represents the 3-link staple terms and $R_\mu(x)$ represents the 5-link rectangle terms in Eq.~(\ref{eq:over-improvement}).
We restrict the over-improvement term such that $\epsilon \in \left[-\frac{5}{2},1\right]$ to ensure
\begin{equation}
    \frac{5-2\epsilon}{3} \geqslant 0 \,,
\end{equation}
and
\begin{equation}
    \frac{\epsilon-1}{12u^2_0} \leqslant 0 \,.
\end{equation}
Then, the minima of the real component of $L^1\left[\Sigma_\mu(x)\right]$ for a given $\alpha$ occur when all 6 terms contributing to $S_\mu(x)$ have nontrivial phase (and hence a real component equal to $-0.5$). 
Replacing any term with the identity necessarily increases the real component of $S_\mu(x)$.
Hence, the values of $S_\mu(x)$ for which the real component of $L^1$ is minimized are given by
\begin{align}
  S_\mu(x) &= ne^{+i\frac{2\pi}{3}} + (6-n)e^{-i\frac{2\pi}{3}} \nn \\
  &= 6\cos\left(\frac{2\pi}{3}\right) + i(2n-6)\sin\left(\frac{2\pi}{3}\right) \nn \\
  &= -3 + i(2n-6)\sin\left(\frac{2\pi}{3}\right)
\end{align}
where $n \in \left[0,6\right] \subset \Z$.
As we are not concerned with the imaginary component, for simplicity and without loss of generality we take $n=3$ for which the imaginary component of $S_\mu(x)$ vanishes.
Evaluating $S_\mu(x)$ for $n=3$, we have $S_\mu(x)=-3$.

On the other hand, as the factor in front of $R_\mu(x)$ is negative, the minima of $L^1$ for a given $\alpha$ occur when the real component of $R_\mu(x)$ is maximized.
This occurs when all 18 loops contributing to $R_\mu(x)$ are the identity with real component equal to 1.
Replacing any term with one which has nontrivial phase (and hence real component equal to $-0.5$) necessarily reduces the real component of $R_\mu(x)$.
Hence, the minima of $L^1$ must occur when $R_\mu(x)=18$.

Substituting into equation (\ref{eq:L_overimp_explicit}) we have
\begin{equation}
    L^1=(1-\alpha) + \frac{\alpha}{6}\left\{\left(\frac{5-2\epsilon}{3}\right)(-3) + \left(\frac{\epsilon-1}{12u^2_0}\right)(18)\right\}\,.
    \label{eq:L_overimp_explicit_minre}
\end{equation}
Recalling that we require $\re{L^1} < 0$, and simplifying we have
\begin{align}
    0 &> (1-\alpha) + \frac{\alpha}{6}\left\{\left(\frac{5-2\epsilon}{3}\right)(-3) + \left(\frac{\epsilon-1}{12u^2_0}\right)(18)\right\} \nn \\
    0 &> 1-\alpha + \alpha \left\{\frac{2\epsilon-5 + \frac{3}{2}\left(\frac{\epsilon-1}{u^2_0}\right)}{6}\right\} \nn \\ 
    -1 &> \alpha \left\{\frac{2\epsilon-11 + \frac{3}{2}\left(\frac{\epsilon-1}{u^2_0}\right)}{6} \right\} \nn \\ 
    \alpha &> \frac{-6}{2\epsilon - 11 + \frac{3}{2}\left(\frac{\epsilon-1}{u_0^2}\right)}\,.
\end{align}

Let $\Sigma_\mu^\text{min}(x)$ denote a staples term which minimizes the real component of $L^1$ for $Z_\mu(x) = \I$.
Then, for $Z_\mu(x) = e^{\pm i\frac{2\pi}{3}}$ the minima of the real component of $L^1$ occur at $e^{\mp i\frac{2\pi}{3}}\Sigma_\mu^\text{min}(x)$, and the same derivation follows.

\bibliographystyle{apsrev4-1}
\bibliography{Bibliography.bib}

\end{document}